# Non-adiabatic phonon renormalization in metallic versus insulating rutile oxides

**Reshma Kumawat[1], Shubham Farswan, Simranjeet Kaur, Kaushik Sen[1]***

[1]Department of Physics, Indian Institute of Technology Delhi, Hauz Khas, New Delhi 110016, India

*Email: kaushik.sen@physics.iitd.ac.in

We present a comparative Raman scattering study of metallic rutile oxides ($RuO_2$ and $IrO_2$) and insulating rutiles ($TiO_2$ and $SnO_2$). Temperature-dependent Raman spectra reveal that metallic compounds exhibit pronounced phonon frequency hardening ($\omega(11\,K) - \omega(300\,K) = \Delta\omega \approx 6 - 10$ cm$^{-1}$), whereas the insulating rutiles show only modest hardening $\Delta\omega \approx 1 - 3$ cm$^{-1}$. In contrast, the linewidth changes $\Delta\Gamma \approx 1 - 7$ cm$^{-1}$ do not display a systematic metallic-insulating classification. Fits with the conventional Klemens anharmonic decay model reproduce the overall temperature trends but yield inconsistent anharmonic parameters for the metallic compounds when benchmarked against insulating rutile analogues. A modified Klemens framework, incorporating an additional $T^2$ correction to the phonon frequency arising from the electronic contribution to the phonon self-energy, quantitatively accounts for the enhanced renormalization observed in metallic systems. These results establish finite non-adiabatic electron-phonon coupling in metallic rutiles and demonstrate that phonon renormalization can be identified even in the absence of observable Fano asymmetry in the phonon lineshapes.

## I. INTRODUCTION

Rutile $RuO_2$ has been an intense research topic due to reports of unconventional magnetic behaviour, including proposals of so-called '*altermagnetism*' [1,2]. However, its magnetic ground state remains debated. While polarized neutron diffraction and resonant x-ray scattering experiments have reported antiferromagnetic order with a magnetic moment of $\sim 0.05\,\mu_B$/Ru [3,4], muon spin rotation and bulk-sensitive optical measurements indicate a nonmagnetic metallic ground state [5–7]. In particular, optical conductivity measurements are consistent with itinerant charge carriers and a conventional metallic electronic structure [7]. These contrasting reports position $RuO_2$ as a prototypical metallic rutile oxide in which the interplay between electronic and lattice degrees of freedom warrants careful examination.

In insulating oxides, the temperature dependence of optical phonons is generally governed by anharmonic phonon-phonon interactions. The conventional Klemens model, describing symmetric decay of an optical phonon into two acoustic phonons, successfully captures both the frequency softening and linewidth broadening with increasing temperature [8]. In metallic systems, however, the presence of itinerant carriers introduces an additional channel: the electronic contribution to the phonon self-energy through electron-phonon coupling. This additional contribution can alter the temperature evolution of the phonon frequency in a manner not captured by purely anharmonic phonon-phonon interactions.

To isolate and quantify this electronic contribution, we perform a comparative Raman scattering study of isostructural rutile oxides: metallic $RuO_2$ and $IrO_2$, and insulating $TiO_2$ and $SnO_2$. The shared tetragonal rutile structure provides a controlled platform in which phonon decay phase space is comparable, allowing deviations from purely anharmonic behavior to be identified systematically. By analyzing the temperature evolution of the Raman-active $E_g$ and $A_{1g}$ modes, we demonstrate that while insulating rutiles are well described by the conventional Klemens framework, the metallic compounds exhibit enhanced frequency renormalization that cannot be accounted by phonon-phonon interactions alone.

To address this discrepancy, we introduce a phenomenological extension of the Klemens description to metallic systems by incorporating the electronic contribution to the phonon self-energy. A systematic comparison with insulating rutiles establishes the purely anharmonic baseline, against which the enhanced frequency renormalization in the metallic compounds is identified. The modified framework quantitatively accounts for this additional electronic contribution while preserving the conventional behavior in the insulating counterparts. Importantly, this analysis demonstrates that finite electron-phonon coupling can be inferred from temperature-dependent phonon renormalization even in the absence of Fano asymmetry in the Raman lineshape. The present work therefore provides a systematic route for separating anharmonic and electronic effects in the lattice dynamics of metallic oxides.

## II. EXPERIMENT

### A. Sample preparation

Polycrystalline $RuO_2$ (Sigma-Aldrich, 99.9%), $TiO_2$ (Thermo Fisher Scientific, 99.9%), $SnO_2$(Tokyo Chemical Industry, 99.9%), and $IrO_2$ (Otto Chemie Pvt. Ltd, 99%) were commercially procured. While $RuO_2$ and $SnO_2$ were obtained entirely in the rutile phase, the as-received $TiO_2$ powder contained both rutile and anatase phases and $IrO_2$ contained hydrous phase. To achieve phase-pure rutile $TiO_2$ and $IrO_2$, we annealed the powder at 1080 °C for 12 hours and at 1000 °C for 6 hours, respectively. For Raman scattering measurements, solid-state pellets were prepared from powder samples.

### B. X-ray diffraction

Symmetric $\theta - 2\theta$ x-ray diffraction (XRD) measurements on powder samples were performed using a Malvern Panalytical Empyrean Series 3 XRD diffractometer (Cu $K_\alpha$ radiation $\lambda = 1.54$ Å) in the range of $20° \leq 2\theta \leq 90°$ with a step size of $2\theta = 0.04°$ at room temperature. Rietveld refinement of the powder patterns was performed using the FullProf suite to extract the lattice parameters and atomic coordinates.

### C. Magnetization

Magnetic dc susceptibility ($\chi$) was measured using a Quantum Design MPMS. Approximately 15 mg of $RuO_2$ was sealed in Teflon tape and mounted in plastic straws perforated to remove trapped oxygen. Field-cooled magnetization was recorded on warming from 2 to 380 K after cooling the samples to 2 K in an applied field of 0.1 T.

### D. Raman scattering

Raman measurements were performed using a home-built setup based on a Horiba iHR550 spectrometer. Temperature-dependent spectra (10-300 K) were collected with an ultra-low-vibration microscopy cryostat (ColdEdge Technologies). Excitation was provided by a He-Ne laser ($\lambda = 632.8$ nm) focused through a $10\times$ long-working-distance objective (NA $= 0.28$), and scattered light was collected in backscattering geometry. A grating of 1200 grooves/mm provided a spectral resolution of $\sim$2.5 cm$^{-1}$. Two notch filters suppressed the Rayleigh line, giving access down to 80 cm$^{-1}$. The laser power was kept below 0.9 mW to avoid local heating.

The phonon modes of all the rutile oxides were fitted with Lorentzian profiles. We also confirmed that the phonon modes of $RuO_2$/$IrO_2$ do not exhibit any Fano asymmetry. Fits using Fano profiles yielded asymmetry parameters that were nearly zero, effectively identical to Lorentzian lineshapes.

## III. RESULTS AND DISCUSSION

### A. Structural characterization

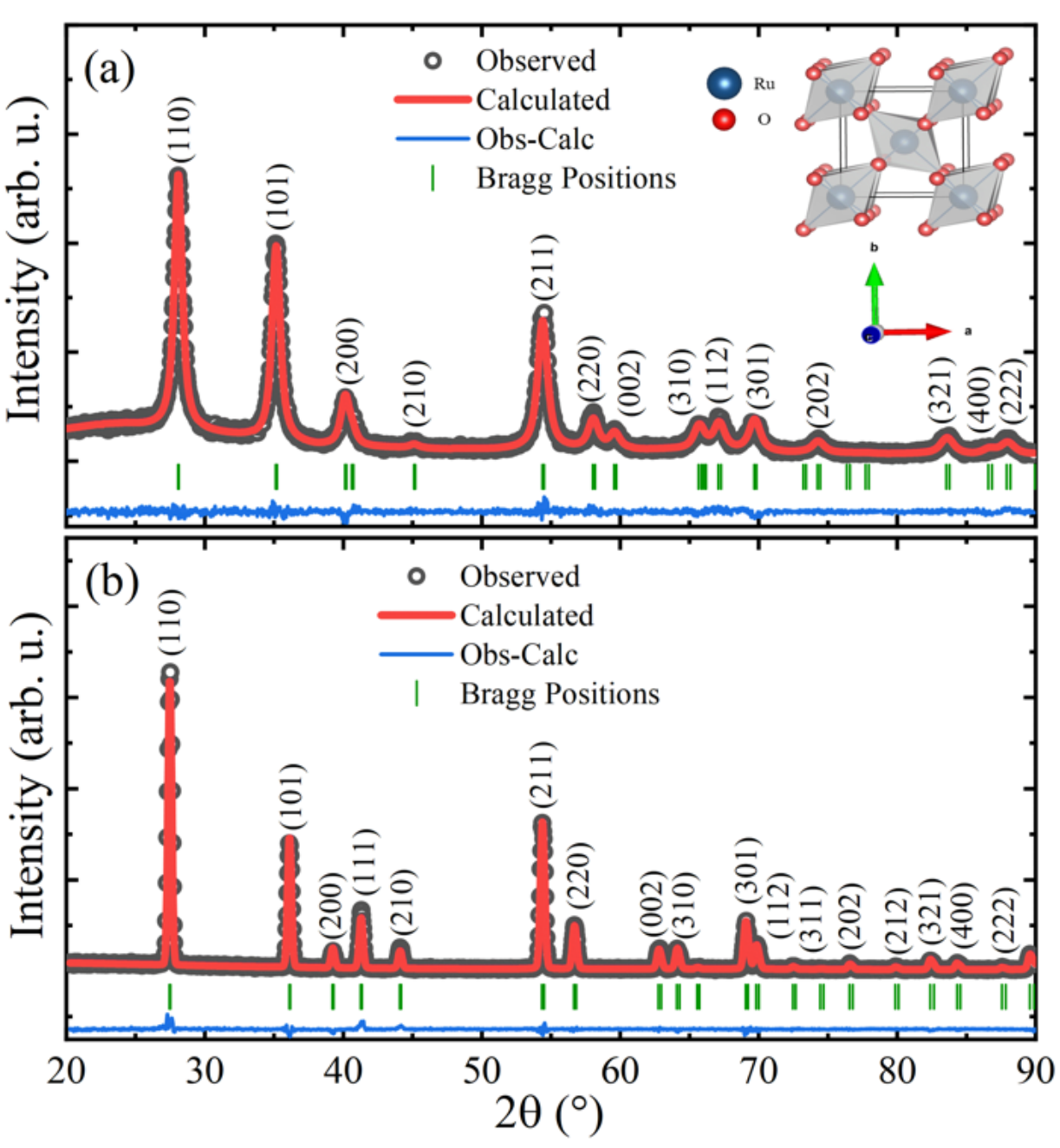

FIG. 1. Symmetric x-ray diffraction patterns (symbols in black) along with the corresponding Rietveld refinements (solid lines in red) for (a) $RuO_2$ and (b) $TiO_2$. The blue line at the bottom shows the difference between observed and simulated data. The green vertical dashes show the positions of the Bragg peaks. The inset in (a) shows the rutile unit cell.

FIG. 1(a) and 1(b) show representative powder XRD patterns of $RuO_2$ and $TiO_2$ at ambient conditions, respectively. The observed Bragg peaks agree well with the reported rutile structures of $RuO_2$ [9] and $TiO_2$ [10]. The narrower peak widths in $TiO_2$ indicate higher crystallinity and a larger average crystallite size (~84 nm) compared to $RuO_2$ (~34 nm), as estimated from the Scherrer method [11]. Rietveld refinements based on the rutile structure (space group $P4_2/mnm$) yielded good fits, with $\chi^2$ values of 1.25 for $RuO_2$ and 1.67 for $TiO_2$. The refined lattice parameters are summarized in TABLE I and closely match reported values, confirming the high quality of the samples. Note that the crystal unit-cell volumes ($V_{\mathrm{UC}} = a^2c$) of $RuO_2$ (62.5 Å$^3$) and $TiO_2$ (62.2 Å$^3$) are nearly equal. The inset of FIG. 1(a) shows the refined $RuO_2$ unit cell obtained using VESTA, consisting of Ru atoms (blue) octahedrally coordinated by O atoms (red) in corner- and edge-sharing $RuO_6$ octahedra (gray). The same crystallographic motif applies to $TiO_2$.

TABLE I. Refined lattice parameters of $RuO_2$ and $TiO_2$ (space group $P4_2/mnm$) compared with reported values.

| | $RuO_2$ | | $TiO_2$ | |
|---|---|---|---|---|
| Lattice parameters | Reported [9] | Refined | Reported [10] | Refined |
| $a$ (Å) | 4.49 | 4.49 | 4.58 | 4.59 |
| $c$ (Å) | 3.10 | 3.10 | 2.95 | 2.95 |

## B. Magnetic Properties

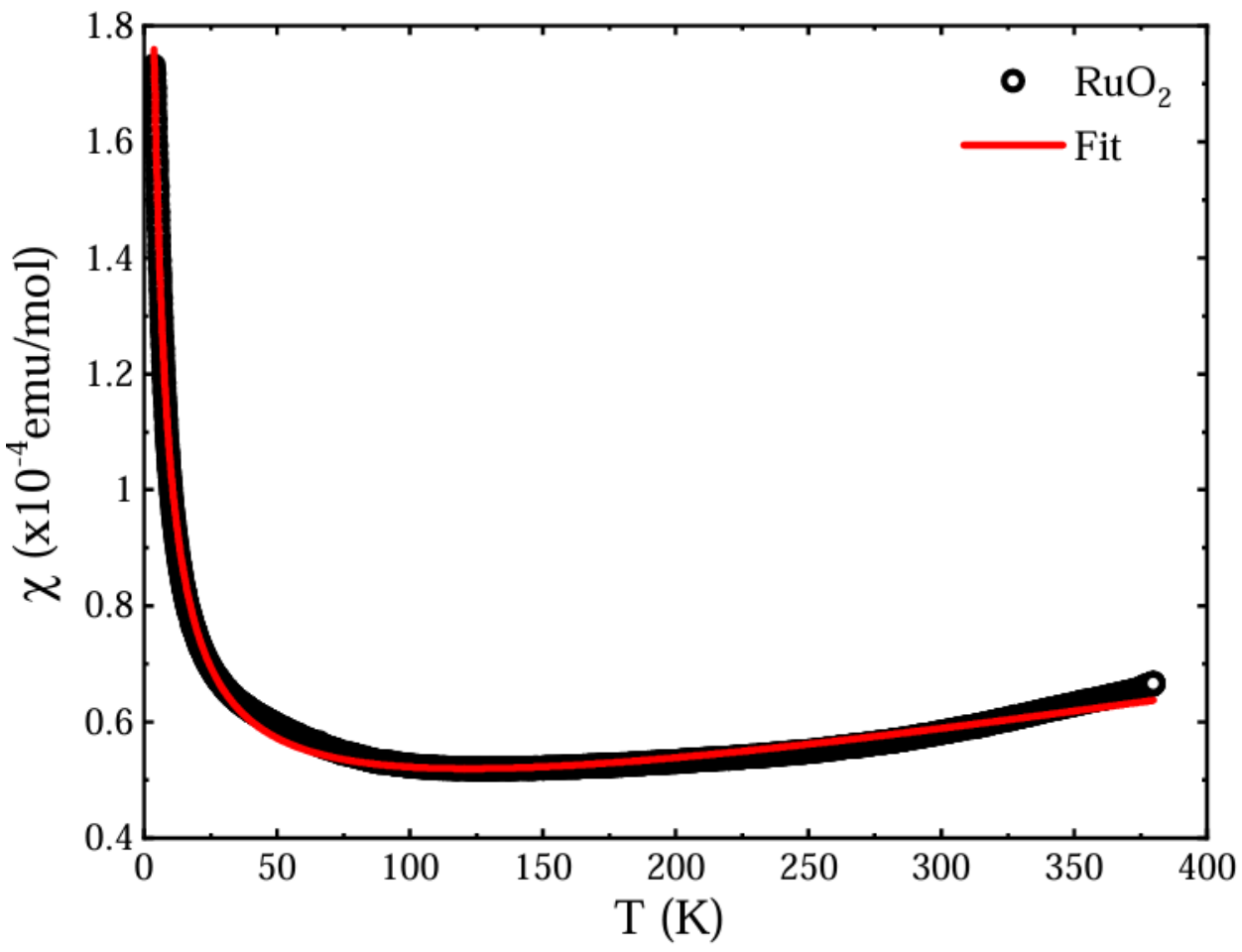

FIG. 2. DC magnetic susceptibility $\chi(T)$ for $RuO_2$, measured under $\mu_0 H = 0.1$T. The data is fitted with a model including a $T^2$ contribution from paramagnetic metal at high temperature and a Curie-like impurity term at low temperature.

FIG. 2 shows the field-cooled dc magnetic susceptibility ($\chi$) as a function of temperature ($T$) of polycrystalline $RuO_2$, measured under an applied field of 0.1 T. The data establish the metallic magnetic response of $RuO_2$ and provide context for its non-adiabatic phonon renormalization discussed below.

For a paramagnet, one expects $\chi$ to follow Curie law, i.e. $\chi \propto 1/T$; however, a clear distinction arises for $RuO_2$. At the low temperature, the sharp rise of $\chi$ towards low temperature is indeed the signature of paramagnetic Curie susceptibility. But this is contrasted with the increase of $\chi$ with $T$ from around 100 K onwards. Our observation is exactly the same as reported for $RuO_2$ in Ref. [7]. Such $\chi(T)$ behavior is typical of $4d$ and $5d$ transition metals, as discussed by Misawa *et al.* [12]. For non-interacting electrons (itinerant electron gas), one expects Pauli paramagnetism, which is temperature independent and defined at $T = 0$ K. Electron-electron interactions introduce a temperature-dependent contribution. In particular, for Fermi-liquid interactions, the susceptibility follows $T^2$ dependence in the intermediate temperature range, leading to an increase in $\chi$ up to temperatures on the order of the Fermi temperature ($T^*$) [12].

For Fermi-liquid-type itinerant carriers, the susceptibility is expected to scale as $\chi(T) \propto T^2\, ln(T/T^*)$ [12]. The logarithmic factor ensures that $\chi(T)$ does not increase indefinitely but instead reaches a maximum near $T^*$. Our data were fitted using the expression

$$\chi(T) = \chi_0 + \frac{C}{T-\theta} - AT^2\, ln\left(\frac{T}{T^*}\right), \quad (1)$$

where $\chi_0$ is the temperature-independent contribution (Pauli plus diamagnetic susceptibility), $C/(T-\theta)$ accounts for the Curie-like term from localized impurity spins, and the last term represents the Fermi-liquid contribution from itinerant carriers [7,12]. A finite Curie-Weiss temperature $\theta$ was included to generalize the model. Because of the limited temperature window, $T^*$ could not be determined directly; instead, we used $T^* \sim 1716$ K obtained by fitting the data over a wide temperature range in Ref. [3] (see Supplementary Information). The resulting parameters, $\chi_0 = (0.43 \pm 0.00058) \times 10^{-4}$ emu/mol, $C = (6.9 \pm 0.03) \times 10^{-4}$ emu K/mol, $\theta = (-1.4 \pm 0.03)$ K, and $A = (8.7 \pm 0.04) \times 10^{-11}$ emu mol$^{-1}$K$^{-2}$ are consistent with the prior report by Wenzel *et al.* [7].

They also reported $T^2$-like temperature dependence of $\chi$ in $RuO_2$, together with optical conductivity evidence for Fermi-liquid dynamics. This independent agreement strongly supports our interpretation that the rise of susceptibility reflects itinerant Fermi-liquid carriers rather than antiferromagnetic correlations reported in Ref. [3]. Further, recent μSR studies [5,6] confirm the absence of static or fluctuating magnetism in both bulk and thin-film $RuO_2$, reinforcing that the observed susceptibility increase originates from itinerant Fermi-liquid carriers.

In summary, these results establish that polycrystalline $RuO_2$ behaves as an itinerant paramagnet with Fermi-liquid interactions persisting up to the highest measured temperatures.

### C. Phonon modes and electronic screening in metallic and insulating rutile oxides

Rutile $RuO_2$ (metallic representative) and $TiO_2$ (insulating representative) belong to the space group of $P4_2/mnm$ and the point group of $D_{4h}$. Group theory analysis at Γ (zone-center) yields four optical phonon modes of $A_{1g} \oplus B_{1g} \oplus B_{2g} \oplus E_g$. All these four modes are Raman-active.

FIG. 3(a) and 3(b) show representative unpolarized Raman spectra at 11 and 300 K for $RuO_2$ and $TiO_2$, respectively. The corresponding data for metallic $IrO_2$ and insulating $SnO_2$ are shown in SI. In the case of $RuO_2$, three modes from low energy to high energy are assigned as $E_g$, $A_{1g}$, and $B_{2g}$ along the line of the first Raman scattering results on single crystals of $RuO_2$ [13]. Note that the $B_{1g}$ mode reported at 97 cm$^{-1}$ was not observed in our measurements for two reasons: it lies below our experimental cutoff of 100 cm$^{-1}$, and it is inherently weak in intensity. In $TiO_2$, we detected four modes: first-order $B_{1g}$, $E_g$, and $A_{1g}$ modes, along with a broad higher-order mode, consistent with the established Raman scattering study on single crystalline $TiO_2$ [14]. Our data from polycrystalline $TiO_2$ closely reproduce the reported Raman spectra, which was measured only down to 100 K [15]. The $B_{2g}$ mode could not be resolved owing to its inherently weak Raman scattering cross-section [15]. All Raman-active modes arise from the vibrations of oxygen ions relative to the nearly static Ru/Ti ions. In the $B_{1g}$, $A_{1g}$, and $B_{2g}$ modes, the oxygen atoms vibrate perpendicular to the c-axis, whereas for the $E_g$ mode, they vibrate along the c-axis [9,15]. In both systems, the phonon modes display a characteristic Lorentzian lineshape.

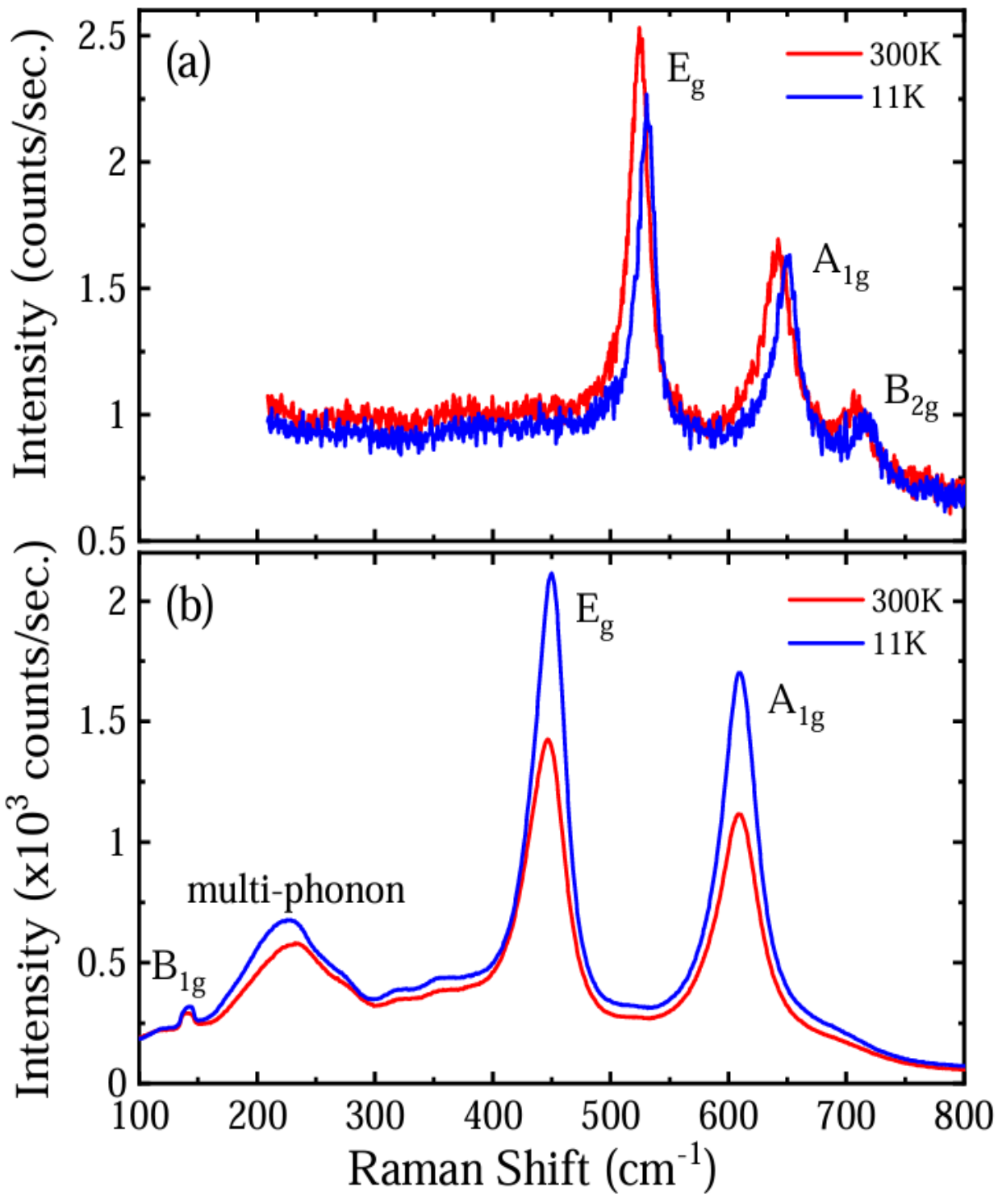

FIG. 3. Unpolarized Raman spectra at 11 and 300 K for (a) $RuO_2$ and (b) $TiO_2$, respectively.

The intensities of the $E_g$ and $A_{1g}$ modes in $RuO_2$ are much weaker, by nearly three orders of magnitude, than those of $TiO_2$ under identical conditions. Such a suppression of phonon intensity is characteristic of metals and originates from adiabatic electronic screening, i.e., conduction electrons respond much faster than the ionic motion and thereby screen the change in polarizability induced by the phonon. The Raman intensity of a mode is determined by the Raman tensor $R$, with $I \propto |\mathbf{e_i}. R. \mathbf{e_s}|^2$, where $\mathbf{e_i}$ and $\mathbf{e_s}$ are the polarization vectors of the incident and scattered lights,

respectively [16]. Since the tensor elements are proportional to $\partial\alpha/\partial Q$, the derivative of the electronic polarizability with respect to the phonon coordinate, strong electronic screening reduces $\partial\alpha/\partial Q$ and hence the Raman intensity. In addition, the penetration depth of the 633 nm excitation laser is drastically smaller in metallic $RuO_2$ compared to $TiO_2$, leading to a significantly reduced effective scattering volume. These explain the stark intensity contrast between $RuO_2$ and $TiO_2$.

In the following, we focus on the two prominent Raman-active modes, $E_g$ and $A_{1g}$, in metallic ($RuO_2$, $IrO_2$) and insulating ($TiO_2$, $SnO_2$) rutile oxides to analyze the temperature evolution of phonon self-energies beyond pure phonon-phonon interactions.

### D. Klemens decay model versus non-adiabatic phonon renormalization in $RuO_2$

FIG. 4 and FIG. 5 show the temperature dependence of the phonon frequency $\omega(T)$ for the $E_g$ and $A_{1g}$ modes, respectively, for metallic ($RuO_2$, $IrO_2$) and insulating ($TiO_2$, $SnO_2$) rutile oxides. FIG. 6 shows the temperature dependence of the corresponding linewidths $\Gamma(T)$. For all compounds studied, the overall behavior looks conventional: the phonon frequency hardens and the linewidth narrows upon cooling. However, the magnitude of phonon renormalization differs markedly between metallic and insulating rutile oxides. In the metallic rutiles $RuO_2$ and $IrO_2$, the $E_g$/$A_{1g}$ phonon modes exhibit pronounced hardening upon cooling from 300 to 11 K, with $\Delta\omega$ = 6-7 $cm^{-1}$ for $RuO_2$ and $\Delta\omega \approx$ 6-10 $cm^{-1}$ for $IrO_2$ (summarized in FIG. 7). In contrast, the insulating rutiles $TiO_2$ and $SnO_2$ show only modest phonon hardening of $\Delta\omega$ = 1-3 $cm^{-1}$. Upon cooling from 300 to 11 K, the linewidth reduction amounts to $\Delta\Gamma \approx$ 1-4 $cm^{-1}$ for $RuO_2$, 4-6 $cm^{-1}$ for $IrO_2$, 4-7 $cm^{-1}$ for $TiO_2$, and 2-4 $cm^{-1}$ for $SnO_2$ (see FIG. 6 and 7). Although the absolute linewidth changes do not follow a single systematic trend across all compounds, the relative balance between $\Delta\omega$ and $\Delta\Gamma$ differs distinctly between metallic and insulating rutiles, as clearly shown in FIG. 7. In $RuO_2$ and $IrO_2$, $\Delta\omega \gtrsim \Delta\Gamma$, whereas in $TiO_2$ and $SnO_2$, $\Delta\Gamma > \Delta\omega$. A similar trend is obtained for other insulating rutile compounds, such as $GeO_2$ and $MgF_2$ by evaluating the reported temperature-dependent Raman data in literature [17,18].

To analyze these data, we applied the standard Klemens anharmonic decay model, in which a zone-center optical phonon decays into two acoustic phonons of equal energy and opposite momenta [19]. The model gives the following linewidth

$$\Gamma(T) = \Gamma_0 + D\left(\frac{\omega_0}{2}\right)\left(1 + \frac{2}{e^{\hbar\omega_0/2k_BT} - 1}\right). \quad (2)$$

The Kramers-Kronig transformation of the above yields the following $\omega(T)$.

$$\omega(T) = \omega_0 - C\left(1 + \frac{2}{e^{\hbar\omega_0/2k_BT} - 1}\right), \quad (3)$$

where, $\Gamma_0$ is the temperature-independent linewidth arising from residual broadening from defects and inhomogeneity, $\omega_0$ is the bare (harmonic) phonon frequency in the absence of anharmonic interactions. $D$ is proportional to three-phonon (cubic) anharmonic coupling coefficient. Within the Klemens framework, the temperature-independent anharmonic coefficient $C$ is determined by the cubic interaction strength $D$, as

$$C = \frac{D\left(\frac{\omega_0}{2}\right)}{2\pi} \ln\left[\left(\frac{\Lambda}{\omega_0}\right)^2 - 1\right], \quad (4)$$

where $\Lambda$ ($\gg \omega_0$) is a high frequency cutoff. A detailed derivation is presented in SI. The solid black lines in FIG. 4-6 show the best fits to $\omega(T)$ and $\Gamma(T)$ using Eqs. (2) and (3). The extracted best-fit parameters $D$ and $C$ are shown in FIG. 8. Although the Klemens expressions reproduce the overall temperature dependence, the extracted parameters reveal an inconsistency within the pure anharmonic framework, i.e., when only phonon-phonon interactions are considered. Because the coefficient $C$ is explicitly determined by the cubic interaction strength $D$, both parameters should reflect the same underlying three-phonon interaction. However, the fitted $D$ values for metallic and insulating rutiles do not exhibit any systematic trend with electronic character. In contrast, $C$ clearly separates into two groups, taking larger values for metallic compounds and smaller values for insulating ones, consistent with the larger $\Delta\omega$ observed in the metallic systems. Meanwhile, the linewidth changes $\Delta\Gamma$, which are governed directly by $D$, do not display a corresponding metallic-insulating separation. These observations indicate that the pure cubic anharmonic model is insufficient to account for the frequency renormalization in metallic rutile oxides. These observations require an additional contribution to the phonon self-energy beyond cubic anharmonicity. In metallic rutile oxides, such a contribution can arise from non-adiabatic corrections associated with the dynamical response of itinerant electrons. Since magnetization and μSR experiments confirm the absence of magnetic order in $RuO_2$ [5,6], a magnetic origin of renormalization can be excluded.

We also note that alternative electronic mechanisms, such as temperature-dependent orbital occupancy or orbital fluctuations, could in principle renormalize phonon frequencies by modifying metal-oxygen covalency. However, our Raman data show a smooth $\omega(T)$ without mode splitting, new modes, or a clear kink. Further, $\Delta\Gamma$ does not exhibit a systematic enhancement in the metallic compounds

Therefore, the most natural mechanism is non-adiabatic electron-phonon coupling, which modifies the phonon self-energy. This motivates a phenomenological modification of the Klemens model, which we develop in the next section.

### E. Modification of Klemens model

For a zone-center optical phonon mode with bare frequency $\omega_0$, the normalized free phonon propagator is written as [20].

$$[\tilde{D}^R(\omega)]^{-1} = (\omega + i0^+)^2 - {\omega_0}^2. \qquad (5)$$

Now, considering the interaction of phonons with other degrees of freedom, Dyson's equation [20] modify the phonon propagator as

$$[\tilde{D}^R(\omega)]^{-1} = (\omega + i0^+)^2 - {\omega_0}^2 - 2\omega_0 \Pi(\omega, T), \qquad (6)$$

here, $\Pi(\omega, T)$ is the complex phonon self-energy, which includes both the normalized phonon frequency (real part) and phonon linewidth (imaginary part).

The pole expansion gives the frequency shift $\Delta\omega\ (\omega(T) - \omega_0)$ and the absolute linewidth as follows [21]

$$\Delta\omega(T) = \Pi'(\omega_0, T), \qquad (7)$$

$$\Gamma(T) = -2\Pi''(\omega_0, T). \qquad (8)$$

Symmetric Klemens decay model gives the anharmonic phonon self-energy, which involves a zone center phonon $(0, \omega_0)$ that decays into two acoustic phonons of equal energy and opposite momenta [19].

$$(0, \omega_0) \rightarrow (\vec{q}, \omega_q) + (-\vec{q}, \omega_{-q}). \qquad (9)$$

As detailed in SI, the phonon linewidth as per symmetric Klemens model reads:

$$\Gamma(T) = \Gamma_0 + D\left(\frac{\omega_0}{2}\right)\left[1 + 2n\left(\frac{\omega_0}{2}\right)\right], \qquad (10)$$

where, $n(\omega)$ is the population of phonons, $\Gamma_0$ is the temperature-independent residual of the linewidth and $D(> 0)$ represents the anharmonic coupling constant from three-phonon processes. Note that this is basically Eq. (2).

We obtained temperature dependence of phonon frequency $(\omega(T))$ from the Kramers-Kronig transformation of $\Gamma(T)$, $\omega(T)$ reads:

$$\omega(T) = \omega_0 - C\left[1 + 2n\left(\frac{\omega_0}{2}\right)\right], \qquad (11)$$

where $C > 0$ is the anharmonic coupling constant, which is proportional to $D$, as given in Eq. (4).

In the presence of electron-phonon interaction, the phonon self-energy can be written as

$$\Pi = \Pi_{ph-ph} + \Pi_{e-ph}, \qquad (12)$$

where, $\Pi_{e-ph}$ is the self-energy from the electron-phonon contribution.

In the Raman limit $q \simeq 0$ (momentum transfer is zero), the non-adiabatic electron-phonon (e-ph) contribution to the phonon self-energy, i.e. the finite-frequency electronic correction beyond the adiabatic Born-Oppenheimer treatment, is governed by interband transitions and is given by [22]

$$\Pi_{e-ph}(\omega_0, T) = \frac{2}{\hbar}\frac{1}{N_k} \sum_k \sum_{m \neq n} |g_{mn}(k)|^2 [f_{mk}(T) - f_{nk}(T)] \left[\frac{1}{\Delta\varepsilon_{mnk} - \hbar(\omega_0 + i\eta)} - \frac{1}{\Delta\varepsilon_{mnk}}\right]. \qquad (13)$$

Here, a factor of 2 accounts for spin degeneracy, and the sums run over electron wave vectors $k$ and all the Kohn-Sham states. The Fermi-Dirac occupations are $f_{nk} = f(\varepsilon_{nk},\mathrm{T})$, where $\varepsilon_{nk}$ are the Kohn-Sham energies and $m, n$ are two electronic band indices. $\Delta\varepsilon_{mnk} = \varepsilon_{mk} - \varepsilon_{nk}$. The quantities $g_{mn}(k)$ denote the electron-phonon matrix elements, and $\eta$ is a real positive infinitesimal.

The real and imaginary parts of Eq. (13) are written as

$$\Pi''_{e-ph}(\omega_0, T) = \frac{2}{\hbar}\frac{1}{N_k} \sum_{k,m\neq n} |g_{mn}(k)|^2 [f_{mk}(T) - f_{nk}(T)] \; \delta(\Delta\varepsilon_{mnk} - \hbar\omega_0) \tag{14}$$

$$\Pi'_{e-ph}(\omega_0, T) = \frac{2}{\hbar}\frac{1}{N_k} \sum_{k,m\neq n} |g_{mn}(k)|^2 [f_{mk}(T) - f_{nk}(T)] \left[\mathcal{P}\frac{1}{\Delta\varepsilon_{mnk} - \hbar\omega_0} - \frac{1}{\Delta\varepsilon_{mnk}}\right] \tag{15}$$

The imaginary part $\Pi''_{e-ph}$ in Eq. (14) arises from real electron-hole pair creation and contributes to the homogeneous phonon broadening. Because it contains a Dirac delta function enforcing strict energy conservation, the available phase space is strongly constrained. In good metals ($\mu \simeq E_F \gg \hbar\omega_0/2$), the occupation difference $f_{mk}(T) - f_{nk}(T)$ varies only weakly with temperature, so the transition is effectively Pauli blocked. Consequently, this phase-space mechanism does not produce a pronounced smooth temperature dependence in $\Gamma(T)$ for $RuO_2$/$IrO_2$. Therefore, the Raman linewidths in $RuO_2$/$IrO_2$ can be reliably described by the anharmonic phonon-phonon contribution alone, as formulated in the Klemens model [Eq. (10) or Eq. (2)]. The phonon frequency renormalization is determined by the real part $\Pi'_{e-ph}$ in Eq. (15), which contains a principal-value sum over electronic states. In the good metal regime $\left(\mu \simeq E_F \gg \frac{\hbar\omega_0}{2}\right)$, the temperature dependence arises from thermal smearing of electronic occupations near the Fermi level rather than from threshold effects. Experimentally, we observe a significantly larger frequency shift in metallic $RuO_2$ and $IrO_2$ compared to insulating $TiO_2$ and $SnO_2$, indicating an additional electronic contribution beyond lattice anharmonicity.

By separating the temperature-independent part, Eq. (15) can be written in the form

$$\Delta\omega_{\mathrm{e-ph}}(\omega_0, T) = \int_{-\infty}^{\infty} \mathcal{K}(\varepsilon; \omega_0) f(\varepsilon, T)\, d\varepsilon + \text{T-independent constant.} \tag{16}$$

Where the kernel $\mathcal{K}(\varepsilon; \omega_0)$ collects all $k$-dependent matrix-element and energy denominator terms, and the temperature dependence enters solely through Fermi-Dirac distribution $f(\varepsilon, T)$.

For smooth $\mathcal{K}(\varepsilon; \omega_0)$ near $\varepsilon = 0$, the Sommerfeld expansion [23] yields a leading quadratic temperature correction to the electronic contribution of the phonon self-energy (detailed in SI). Accordingly, frequency shift is expressed as

$$\Delta\omega_{\mathrm{e-ph}}(T) = \Delta\omega_{\mathrm{e-ph}}(0) + KT^2, \tag{17}$$

where $K = \frac{\pi^2}{6} {K_B}^2 \mathcal{K}'(0; \omega_0)$, and $\mathcal{K}'(0; \omega_0)$ denotes the energy derivative of the kernel.

$\Delta\omega_{\mathrm{e-ph}}(0)$ absorbs temperature-independent parts. The detailed calculations are shown in SI. By absorbing the temperature-independent electronic renormalization $\Delta\omega_{\mathrm{e-ph}}(0)$ into an effective bare frequency $\omega_0^*$, we get $\omega(T)$ as

$$\omega(T) = \omega_0^* - C\left[1 + 2\mathrm{n}\left(\frac{\omega_0^*}{2}\right)\right] + KT^2 \tag{18}$$

where, the $KT^2$ term captures the low-temperature electronic correction arising from electron-phonon interactions and the sign of $K$ depends on the electronic bandstructure of the material.

Thus, by replacing $\omega_0$ with $\omega_0^*$ in Eq. (10), the $\Gamma(T)$ for $RuO_2$/$IrO_2$ can be written as

$$\Gamma(T) = \Gamma_0 + D\left(\frac{\omega_0^*}{2}\right)\left[1 + 2n\left(\frac{\omega_0^*}{2}\right)\right] \tag{19}$$

### F. Underscoring non-adiabatic phonon renormalization using modified Klemen's model

The anomalous temperature dependence of the $RuO_2$/$IrO_2$ phonons ($A_{1g}$ and $E_g$) was further analyzed using the modified Klemens model introduced in Eqs. (18) and (19). The best fits are shown by the orange curves in FIG. 4(a)-4(b) and FIG. 5(a)-5(b) for $\omega$ vs. $T$ for $E_g$ and $A_{1g}$, respectively. The extracted parameters are summarized in TABLE II, which compares $C$, $D$, and $\omega_0$ obtained from the standard and modified Klemens model for $RuO_2/IrO_2$. In fitting the metallic $RuO_2/IrO_2$ within the modified Klemens framework, $C$ was fixed to average values obtained from the standard Klemens analysis of insulating rutiles ($TiO_2$, $SnO_2$, $GeO_2$, and $MgF_2$), while $D$ was allowed to vary freely. This choice is justified by the common rutile crystal structure, which ensures comparable phonon decay phase space, while these insulating systems provide reference cases where the temperature evolution is governed purely by phonon-phonon interactions. TABLE II further summarizes the best-fit parameter $K$ for $RuO_2/IrO_2$, which originates from non-adiabatic electron-phonon coupling and is sensitive to the underlying electronic band structure. The inclusion of the additional $T^2$ term within the modified Klemens framework metallic rutiles results in a systematic improvement of the fits, as shown in FIG. 4(a)–4(b) for $E_g$ and 5(a)–5(b) for $A_{1g}$ modes. In the harmonic limit, the phonon frequency is expected to saturate at low temperature. However, the metallic compounds exhibit a continued increase in $\omega(T)$ upon cooling, most pronounced for the $A_{1g}$ mode of $IrO_2$ (FIG. 5(b)). The modified model captures this while retaining physically consistent anharmonic parameters.

As discussed in Sec. E, neither $\Delta\Gamma$ nor the extracted cubic coefficient $D$ displays a systematic metallic-insulating separation. The linewidth data for metallic rutiles in FIG. 6(a) and 6(b) are well described by Eq. (19), where the electron-phonon contribution modifies the harmonic-limit frequency $\omega_0^*$.

It is worth noting that we consistently obtain $D > C$ for the insulating compounds using the standard Klemens model and for the metallic compounds using the modified Klemens model, as shown in FIG. 8. This is consistent with the relation between $C$ and $D$ in Eq. (4). Within the Klemens picture, a phonon of frequency $\omega_0$ decays into two phonons of frequency $\omega_0/2$. Taking the highest phonon frequency in these oxides to be about 800 $cm^{-1}$ [see FIG. 2 in the Supplementary Information for $SnO_2$], the cutoff energy can be estimated as $\Lambda \approx 1600$ $cm^{-1}$. This estimate shows that for phonon modes above 70 $cm^{-1}$, one expects $D > C$, which is indeed satisfied for all the phonon modes studied in the present work.

We note that the conventional Raman signature of electron-phonon coupling is a Fano asymmetry [24]. However, the phonon modes in $RuO_2/IrO_2$ are well described by Lorentzian profiles as mentioned in Sec. III and SI. In general, Fano lineshapes arise when Raman-active phonons couple to a broad electronic continuum. The absence of visible asymmetry therefore suggests that the electronic Raman scattering must be weak, even though the electron-phonon coupling remains finite.

TABLE II. Comparison of fitting parameters for $RuO_2$ and $IrO_2$ obtained from the standard and modified Klemens models (M-Klemens). In the modified model, $C$ was taken from the average of insulating compounds (purely anharmonic references), while $K$ captures the electronic correction to the phonon frequency arising from electron-phonon coupling. Values in brackets indicate the fitting error bars.

| Mode | System | Model | $\omega_0$ ($cm^{-1}$) | $C$ ($cm^{-1}$) | $D$ ($cm^{-1}$) | $K$ ($-\times 10^{-5} cm^{-1}K^{-2}$) |
|---|---|---|---|---|---|---|
| $E_g$ | $RuO_2$ | Klemens | 535.53(16) | 7.76(13) | 2.12(29) | |
| | | M- Klemens | 529.38(10) | 1.45(10) | 2.06(29) | 5.6(1) |
| | $IrO_2$ | Klemens | 571.85(46) | 8.56(44) | 9.20(67) | |
| | | M- Klemens | 564.88(10) | 1.45(10) | 8.97(66) | 5.5(2) |
| $A_{1g}$ | $RuO_2$ | Klemens | 660.03(57) | 13.48(50) | 6.21(102) | |
| | | M-Klemens | 649.21(10) | 2.38(10) | 5.89(100) | 6.6(2) |

| | | | | | |
|---|---|---|---|---|---|
| $IrO_2$ | Klemens | 779.57(207) | 25.12(192) | 11.7(11) | |
| | M- Klemens | 758.43(10) | 2.38(10) | 11.0(11) | 11.5(4) |

## IV. SUMMARY AND CONCLUSION

We performed a comparative Raman scattering study of phonons in metallic rutiles ($RuO_2$, $IrO_2$) and insulating rutiles ($TiO_2$, $SnO_2$). DC magnetic susceptibility data of the representative metallic rutile $RuO_2$ confirms that the compound remains a nonmagnetic itinerant system across the studied temperature range. Phonon modes in metallic rutiles exhibit pronounced hardening upon colling from 300 to 11K, with $\Delta\omega \approx 6 - 10$ cm$^{-1}$. In contrast, the insulating rutile oxides show only modest phonon hardening of $\Delta\omega \approx 1 - 3$cm$^{-1}$. While $\Delta\omega$ differentiate metallic and insulating rutile oxides into two distinct classes, whereas the corresponding linewidth changes $\Delta\Gamma$ do not exhibit the classification.

The Klemens anharmonic phonon decay model reproduces the overall temperature trends but yield unphysical anharmonic constants for metallic rutiles, signaling the breakdown of the purely anharmonic phonon-phonon description. We therefore introduced a modified Klemens framework incorporating an additional $T^2$ correction to the phonon frequency. This term originates from the temperature dependence of the real part of the electron-phonon self-energy, where all thermal variation enters through the Fermi-Dirac distribution function. Expanding the electronic contribution near the Fermi level yields a leading quadratic $T^2$ correction to the phonon frequency. In contrast, the imaginary part of the electronic self-energy remains weakly temperature dependent due to Pauli blocking, so the linewidth is adequately described by the conventional anharmonic phonon-phonon contribution. The analysis establishes sizable electron-phonon coupling in metallic rutiles $RuO_2$ and $IrO_2$.

Our comparative study demonstrates that the absence of Fano asymmetry in a Raman spectrum does not imply negligible electron-phonon coupling, since phonon renormalization can arise from electronic self-energy effects even when the electronic Raman continuum is weak. The modified Klemens framework developed here provides a practical approach for analyzing phonon renormalization in metallic systems.

## ACKNOWLEDGEMENTS

This work is supported by the YSRP Grant from BRNS, INSPIRE Faculty Fellowship from DST, and Core Research Grant from SERB. RK acknowledges her PhD fellowship from IIT Delhi. All authors acknowledge the support from CRF and the Physics department at IIT Delhi. KS and RK acknowledge Siddhartha Lal for valuable discussions on the phenomenological theory presented in this work.

## DATA AVAILABILITY

The data that support the findings of this article are openly available [25].

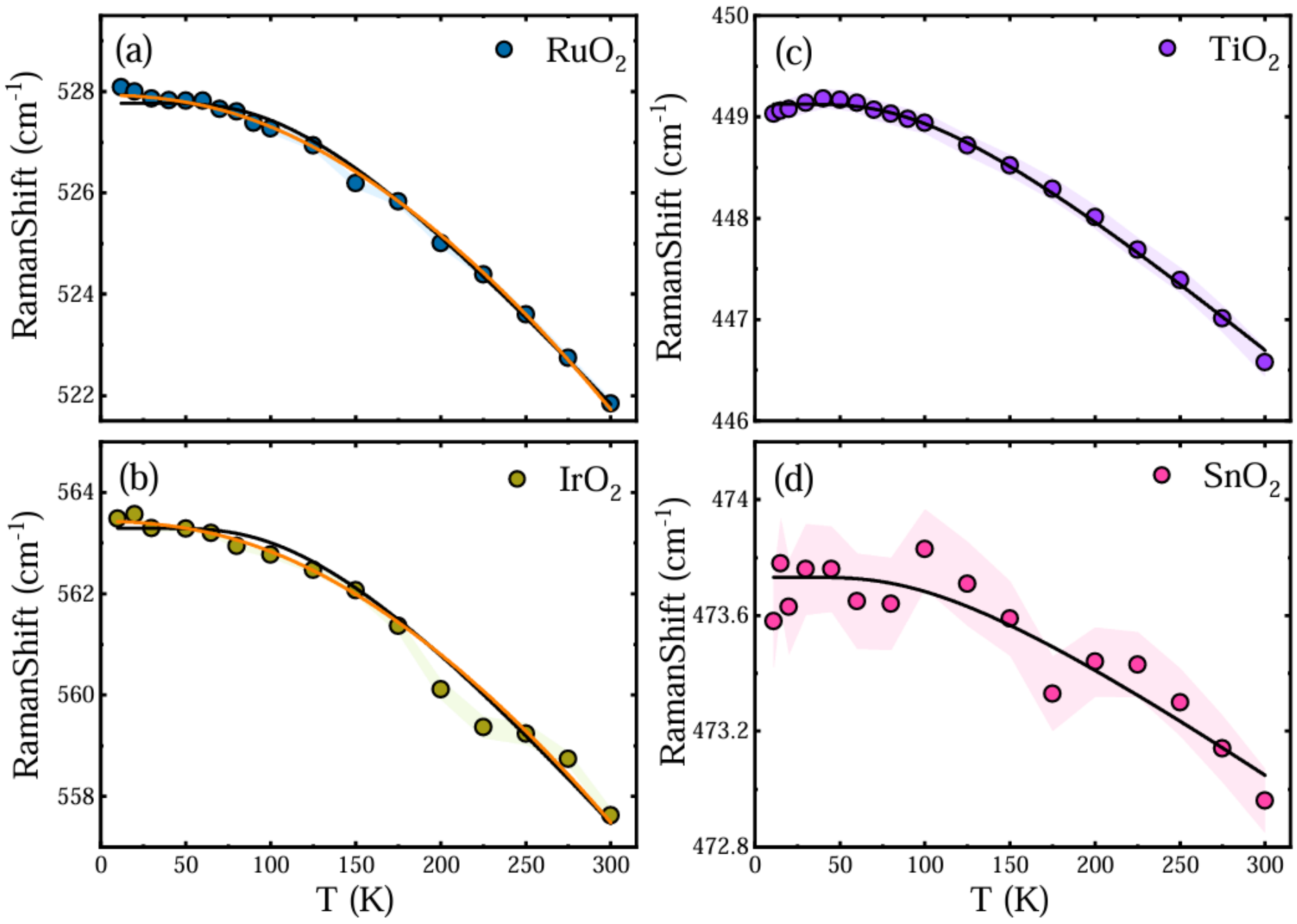

FIG. 4. (a)/(b) and (c)/(d) are the Raman shift as a function of temperature for $E_g$ phonon modes in $RuO_2$/$IrO_2$ and $TiO_2$/$SnO_2$, respectively. The shaded regions indicate the uncertainty range, i.e. the error bars obtained from the nonlinear least-squares Lorentz fitting routine. The solid lines (black) in panels (a)-(d) are the best fits to the data with the standard Klemens decay model, while the solid lines (orange) in panels (a)-(b) are the best fits to the data with the modified Klemens decay model for the metallic $RuO_2$ and $IrO_2$.

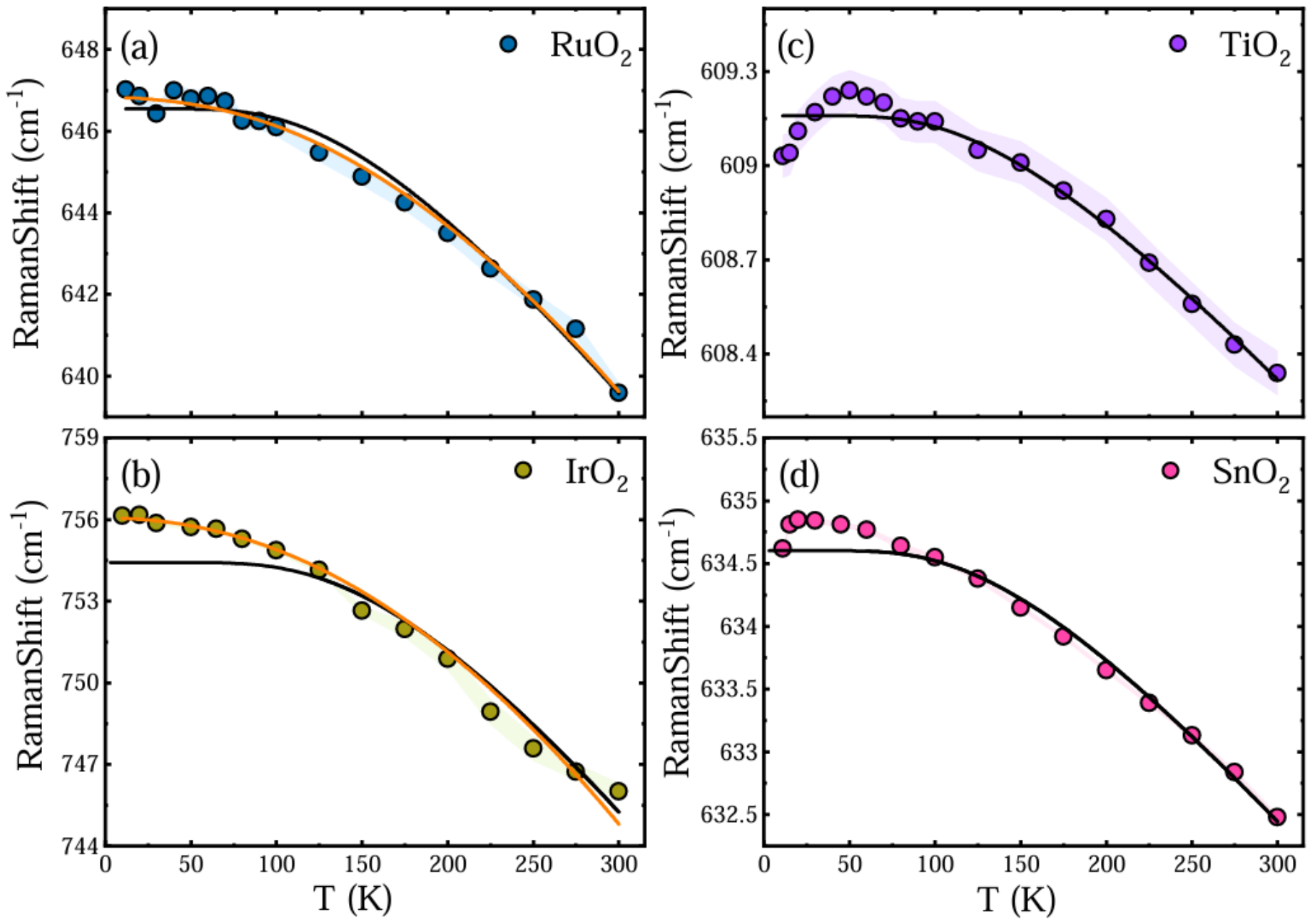

FIG. 5. (a)/(b) and (c)/(d) are the Raman shift as a function of temperature for $A_{1g}$ phonon modes in $RuO_2$/$IrO_2$ and $TiO_2$/$SnO_2$, respectively. The shaded regions indicate the uncertainty range, i.e. the error bars obtained from the nonlinear least-squares Lorentz fitting routine. The solid lines (black) in panels (a)-(d) are the best fits to the data with the standard Klemens decay model, while the solid lines (orange) in panels (a)-(b) are the best fits to the data with the modified Klemens decay model for the metallic $RuO_2$ and $IrO_2$.

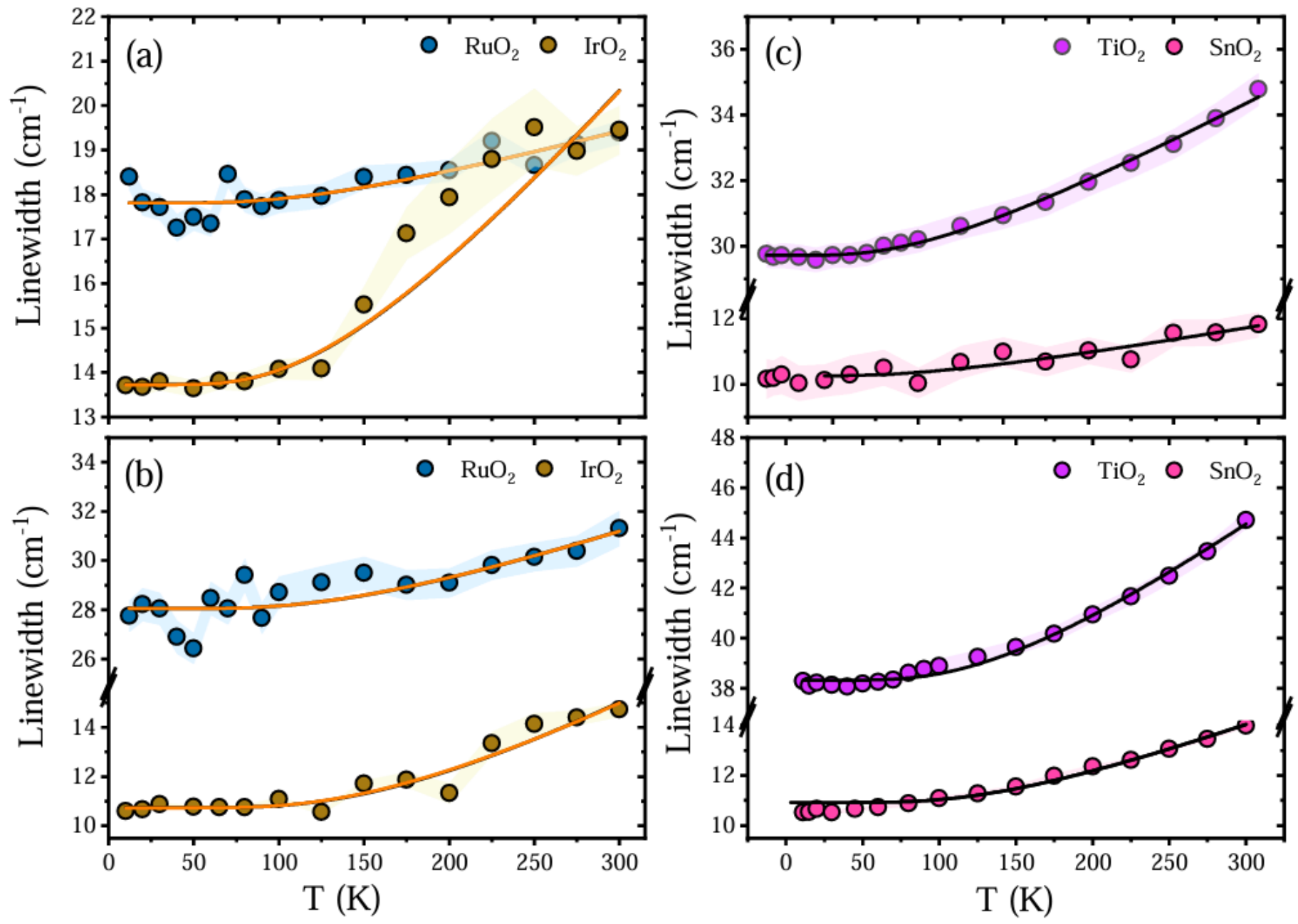

FIG. 6. (a) & (c) and (b) &(d) are linewidth as a function of temperature for $E_g$ and $A_{1g}$ modes of all materials, respectively. The shaded regions indicate the uncertainty range, i.e. the error bars obtained from the nonlinear least-squares Lorentz fitting routine. The solid lines (black) in panels (a)-(d) are the best fits to the data with the standard Klemens decay model, while the solid lines (orange) in panels (a)-(b) are the best fits to the data with the modified Klemens decay model for the metallic $RuO_2$ and $IrO_2$.

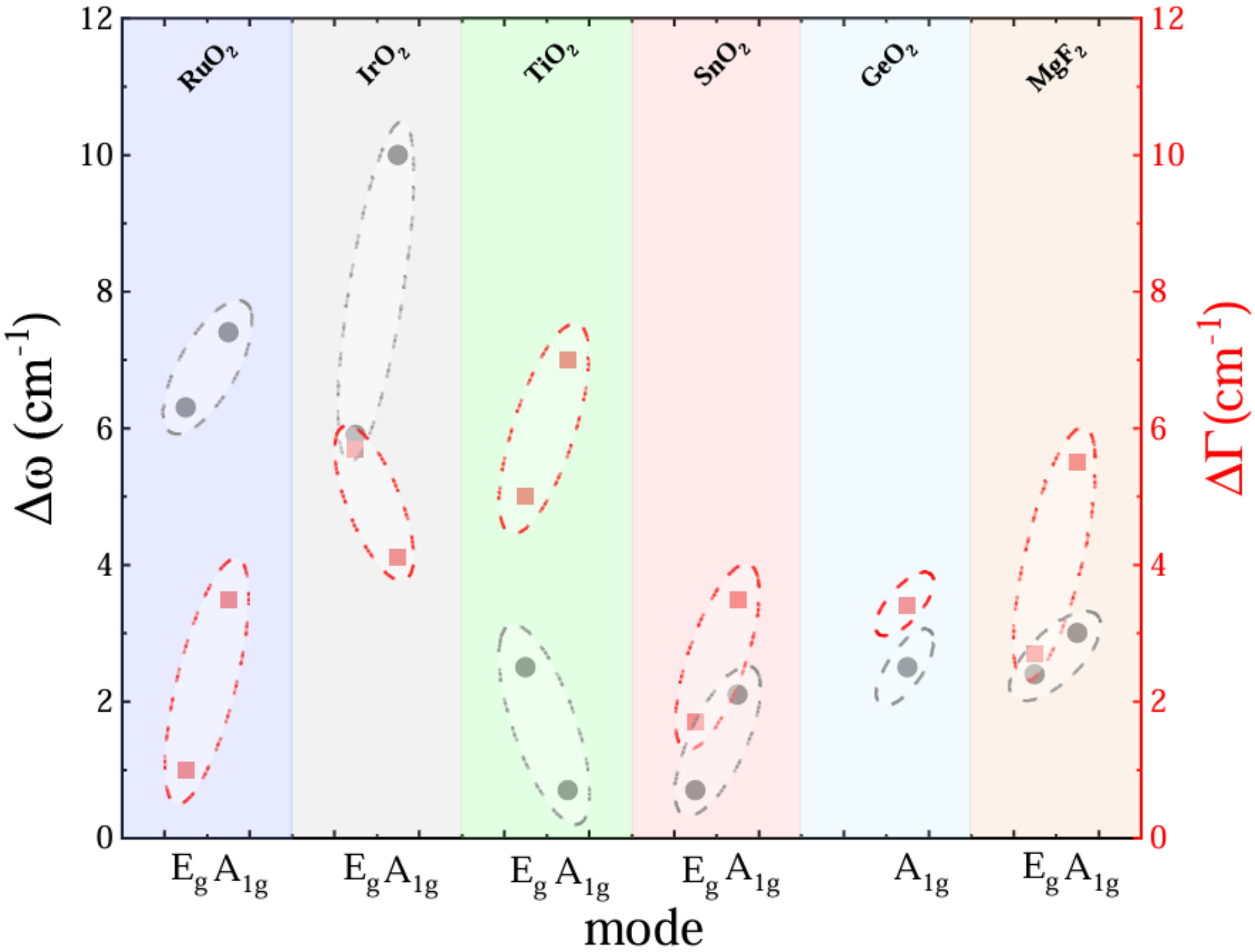

FIG. 7. Comparison of $\Delta\omega(\omega_{11K}-\omega_{300K})$ and $\Delta\Gamma(\Gamma_{11K}-\Gamma_{300K})$ for several isostructural metallic and insulating systems. The data for $GeO_2$ and $MgF_2$ were taken from refs. [17] and [18], respectively.

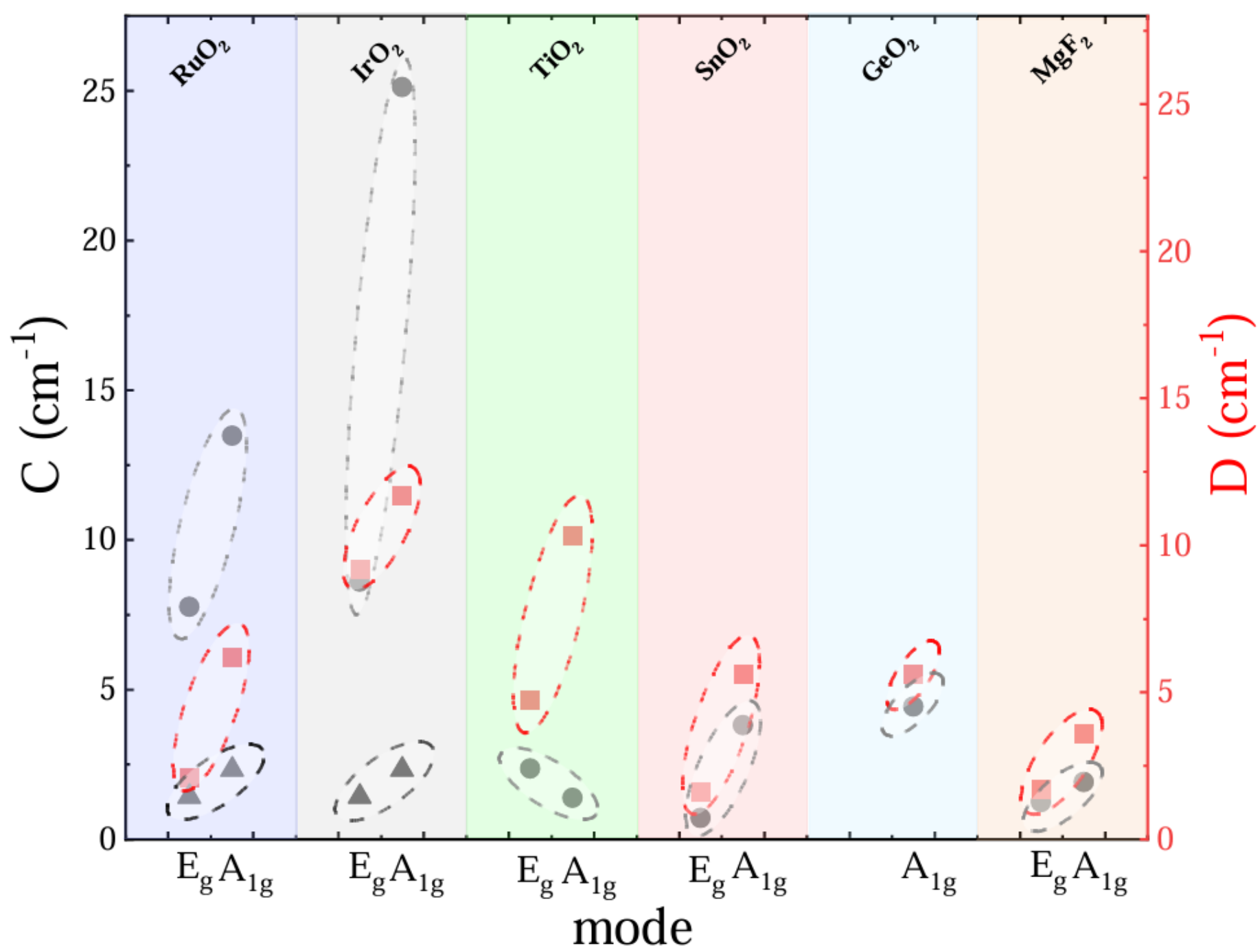

FIG. 8. Comparison of fitting parameters $C$ (gray circles) and $D$ (red squares) of phonon frequency and linewidth with standard Klemens decay model for metallic and insulating rutiles, and $C$ (gray triangles) with the modified Klemens decay model for metallic rutiles.

## References

[1] L. Šmejkal, J. Sinova, and T. Jungwirth, Emerging Research Landscape of Altermagnetism, Phys. Rev. X **12**, 040501 (2022).

[2] L. Šmejkal, J. Sinova, and T. Jungwirth, Beyond Conventional Ferromagnetism and Antiferromagnetism: A Phase with Nonrelativistic Spin and Crystal Rotation Symmetry, Phys. Rev. X **12**, 031042 (2022).

[3] T. Berlijn, P. C. Snijders, O. Delaire, H. D. Zhou, T. A. Maier, H. B. Cao, S. X. Chi, M. Matsuda, Y. Wang, M.R. Koehler, P.R.C. Kent, and H.H. Weitering, Itinerant Antiferromagnetism in $RuO_2$, Phys. Rev. Lett. **118**, 077201 (2017).

[4] Z. H. Zhu, J. Strempfer, R. R. Rao, C. A. Occhialini, J. Pelliciari, Y. Choi, T. Kawaguchi, H. You, J. F. Mitchell, Y. Shao-Horn, and R. Comin, Anomalous Antiferromagnetism in Metallic $RuO_2$ Determined by Resonant X-ray Scattering, Phys. Rev. Lett. **122**, (2019).

[5] P. Keßler, L. Garcia-Gassull, A. Suter, T. Prokscha, Z. Salman, D. Khalyavin, P. Manuel, F. Orlandi, I. I. Mazin, R. Valentí, and S. Moser, Absence of magnetic order in $RuO_2$: insights from μSR spectroscopy and neutron diffraction, Npj Spintron. **2**, 50 (2024).

[6] M. Hiraishi, H. Okabe, A. Koda, R. Kadono, T. Muroi, D. Hirai, and Z. Hiroi, Nonmagnetic Ground State in $RuO_2$ Revealed by Muon Spin Rotation, Phys. Rev. Lett. **132**, 166702 (2024).

[7] M. Wenzel, E. Uykur, S. Rößler, M. Schmidt, O. Janson, A. Tiwari, M. Dressel, and A. A. Tsirlin, Fermi-liquid behavior of nonaltermagnetic $RuO_2$, Phys. Rev. B **111**, L041115 (2025).

[8] P. G. Klemens, Anharmonic decay of optical phonons, Phys. Rev. **148**, 845 (1966).

[9] L. Kiefer, F. Wirth, A. Bertin, P. Becker, L. Bohatý, K. Schmalzl, A. Stunault, J. A. Rodríguez-Velamazan, O. Fabelo, and M. Braden, Crystal structure and absence of magnetic order in single-crystalline $RuO_2$, J. Phys. Condens. Matter **37**, 135801 (2025).

[10] S. Challagulla, K. Tarafder, R. Ganesan, and S. Roy, Structure sensitive photocatalytic reduction of nitroarenes over $TiO_2$, Sci. Rep. **7**, 8783 (2017).

[11] A. L. Patterson, The Scherrer Formula for X-Ray Particle Size Determination, Phys. Rev. **56**, 978 (1939).

[12] S. Misawa and K. Kanematsu, Susceptibility maximum and Fermi-liquid effect in 4d and 5d transition metals, J. Phys. F Met. Phys. **6**, 2119 (1976).

[13] Y. S. Huang and F. H. Pollak, Raman investigation of rutile $RuO_2$, Solid State Commun. **43**, 921 (1982).

[14] S. P. S. Porto, P. A. Fleury, and T. C. Damen, Raman Syectra of $TiO_2$, $MgF_2$, $ZnF_2$, $FeF_2$, and $MnF_2$, Phys. Rev. **154**, 522 (1967).

[15] T. Lan, X. Tang, and B. Fultz, Phonon anharmonicity of rutile $TiO_2$ studied by Raman spectrometry and molecular dynamics simulations, Phys. Rev. B **85**, 094305 (2012).

[16] R. Loudon, The Raman effect in crystals, Adv. Phys. **50**, 813 (2001).

[17] T.P. Mernagh and L.-g. Liu, Temperature dependence of Raman spectra of the quartz- and rutile-types of $GeO_2$, Phys Chem Miner. **24**, 7 (1997).

[18] A. Perakis, E. Sarantopoulou, Y. S. Raptis, and C. Raptis, Temperature dependence of Raman scattering and anharmonicity study of $MgF_2$, Phys. Rev. B **59**, 775 (1999).

[19] K. Sen, Y. Yao, R. Heid, A. Omoumi, F. Hardy, K. Willa, M. Merz, A. A. Haghighirad, and M. Le Tacon, Raman scattering study of lattice and magnetic excitations in CrAs, Phys. Rev. B **100**, 104301 (2019).

[20] A. B. Migdal, Interaction between electrons and lattice vibrations in a normal metal, J. Exptl. Theor. Phys. **34**, 1438 (1958).

[21] I. P. Ipatova and A. V. Subashiev, Long-wave optical-phonon spectrum in metals and heavily doped semiconductors, Zh. Eksp. Teor. Fiz. **66**, 722 (1974).

[22] M. Lazzeri and F. Mauri, Nonadiabatic Kohn anomaly in a doped graphene monolayer, Phys. Rev. Lett. **97**, 266407 (2006).

[23] N. W. Ashcroft and N. D. Mermin, Solid State Physics (Harcourt, Orlando, FL, 1976).

[24] K. Sen, D. Fuchs, R. Heid, K. Kleindienst, K. Wolff, J. Schmalian, and M. Le Tacon, Strange semimetal dynamics in $SrIrO_3$, Nat. Commun. **11**, 1 (2020).

[25] R. Kumawat, S. Farswan, S. Kaur, and K. Sen, Non-adiabatic phonon renormalization in metallic versus insulating rutile oxides, https://drive.google.com/drive/folders/1rhMFxFJJ2E98vcYulFyPki6TJjmCd5tO?usp=sharing

## SUPPLEMENTARY INFORMATION

R. Kumawat *et al.*

### 1. Magnetic Susceptibility

The temperature-dependent magnetic susceptibility of the $RuO_2$ crystal from Ref. [1] was fitted with Eq. (1) as shown in FIG. 1.

$$\chi(T) = \chi_0 + \frac{C}{T-\theta} - AT^2 \ln\left(\frac{T}{T^*}\right). \quad (1)$$

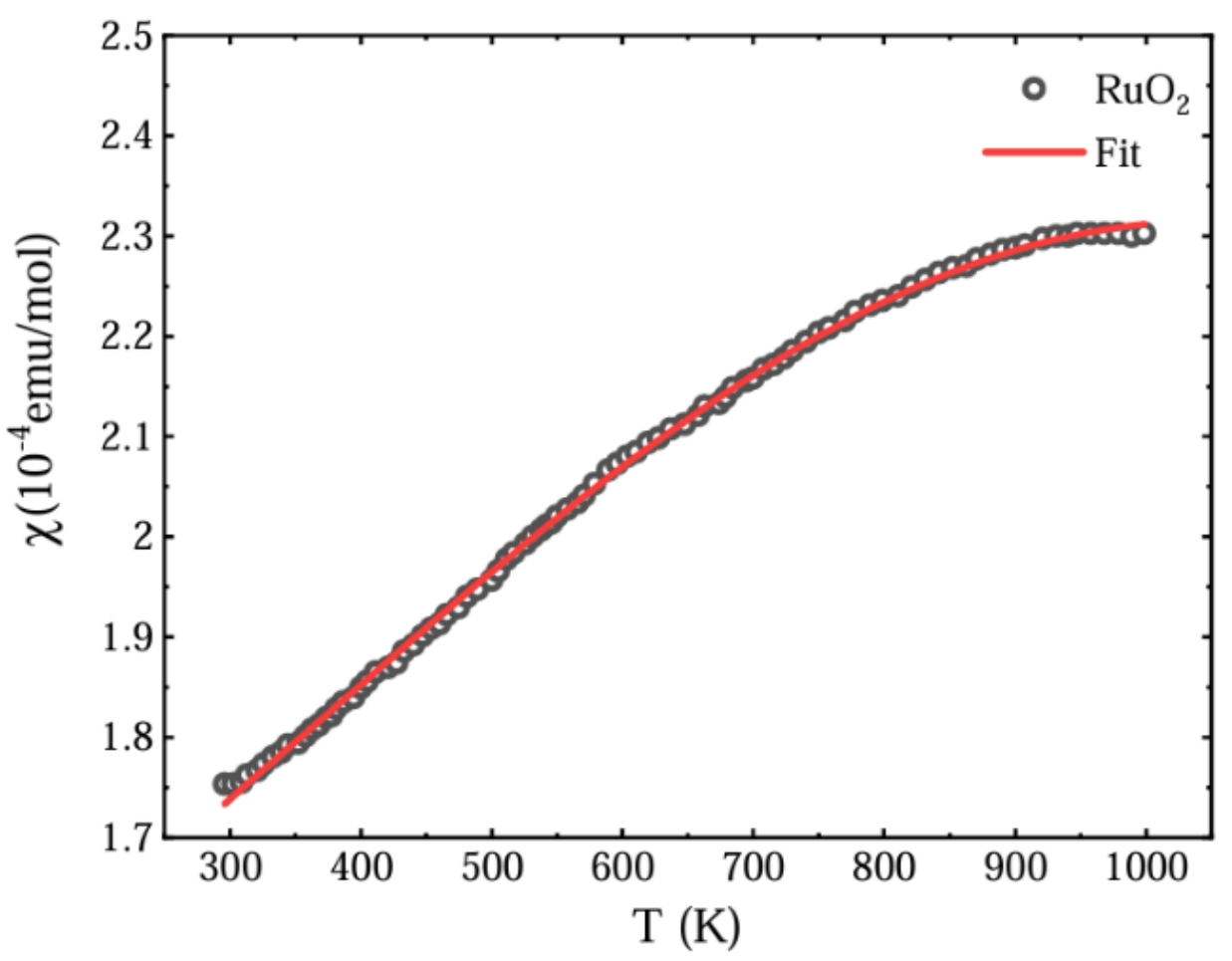

FIG. 1. Magnetic susceptibility $\chi(T)$ for $RuO_2$ as a function of temperature, measured under $\mu_0 H = 2$T.

### 2. Representative Raman spectra of $IrO_2$ and $SnO_2$

In the case of $IrO_2$, three Raman-active modes: $E_g$, $B_{2g}$, and $A_{1g}$ were observed, consistent with the polarization-resolved Raman scattering results on single crystals of rutile $IrO_2$ [2]. The sharp low-frequency $B_{1g}$ mode at 145 $cm^{-1}$ was not detected due to its inherently weak intensity. In $SnO_2$, we identified three modes: the low energy $E_g$ mode and the high-energy $A_{1g}$ and $B_{2g}$ modes, while the weak $B_{1g}$ mode at 123 $cm^{-1}$ could not be resolved [3]. Our data, in close agreement with reported single-crystal spectra, confirm that both $IrO_2$ and $SnO_2$ adopt the rutile structure, with the phonon modes well described by a Lorentzian lineshape.

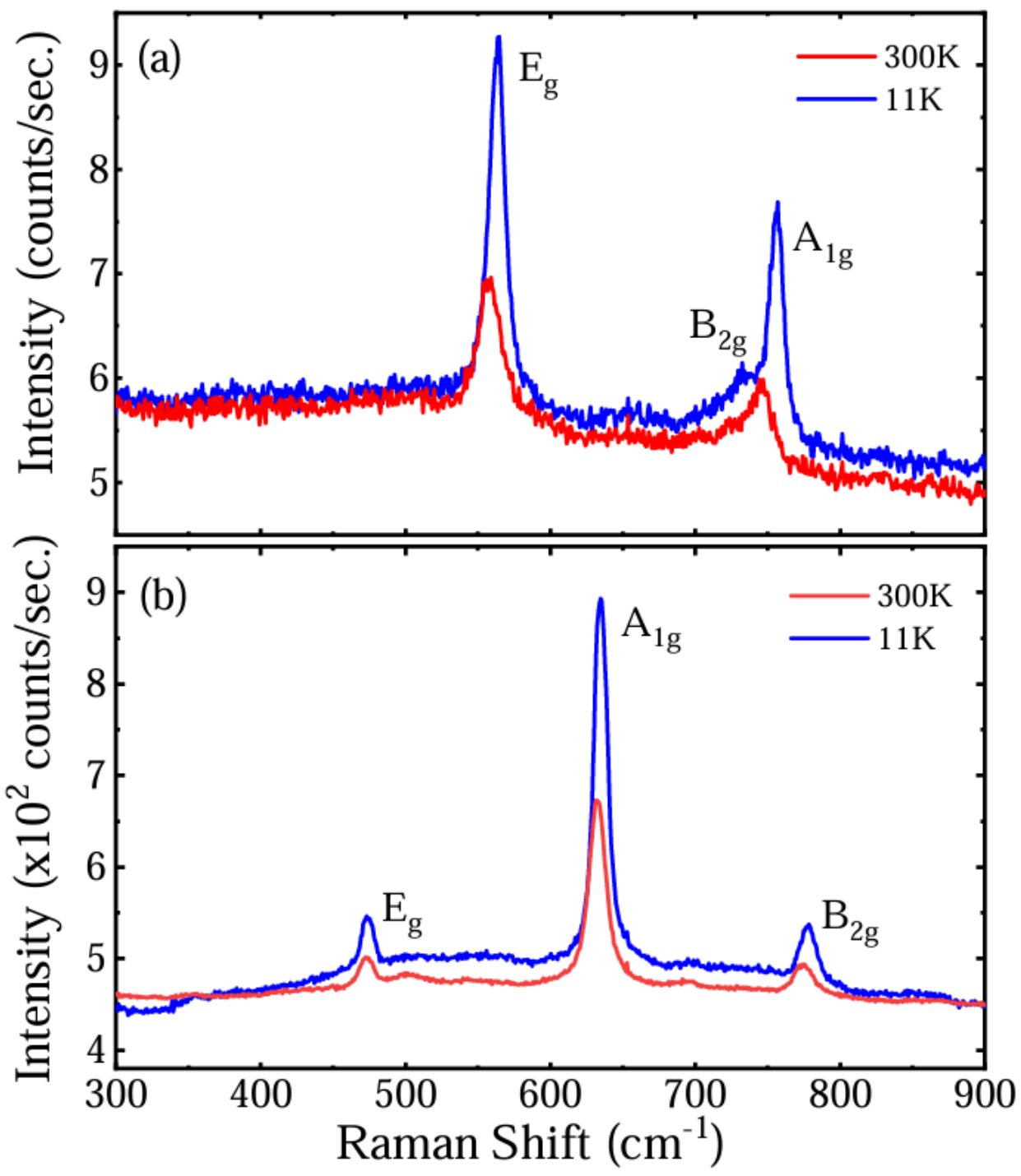

FIG. 2. Unpolarized Raman spectra at 11 and 300 K for (a) $IrO_2$ and (b) $SnO_2$, respectively. The 11K spectra in (b) is vertically shifted for clarity.

## 3. Modification of Klemens model

### 3.1 Klemens Symmetric Phonon Decay Model

The Hamiltonian for quantized lattice vibrations (phonons) is expressed as [4]

$$H_{ph} = \sum_{q} \omega_0 {a^\dagger}_q a_q, \tag{2}$$

where, $a^\dagger(a)$ are creation (annihilation) operators for phonons with momentum $q$ in the Heisenberg picture.

We define the phonon normal coordinate $Q(t)$, which represents the collective displacement operator of the gamma point (zone-center) optical phonon mode with bare frequency $\omega_0$ as

$$Q(t) = \sqrt{\frac{\hbar}{2\omega_0}}\left(a(t) + a^\dagger(t)\right), \tag{3}$$

To understand the quasiparticle renormalization, we define the retarded Green's function associated with the displacement operator of phonons as [4]

$$D^R(t) = -\frac{i}{\hbar}\Theta(\mathrm{t})\langle[Q(t), Q(0)]\rangle, \tag{4}$$

where, $\Theta$ is the Heaviside step function ($\Theta(\mathrm{t}) = 1$ for $t > 0,\ 0$ otherwise). For the harmonic case, the creation and annihilation operators evolve as

$$a(t) = ae^{-i\omega_0 t}, a^\dagger(t) = a^\dagger e^{i\omega_0 t}. \tag{5}$$

For boson, $\left[a, a^\dagger\right] = 1$, which ensures that $D^R(t)$ has a measurable response function via linear response theory.

$$[Q(t), Q(0)] = \frac{\hbar}{2\omega_0}\left(e^{-i\omega_0 t} - e^{i\omega_0 t}\right). \tag{6}$$

For $t > 0$,

$$D^R(t) = -\frac{i}{2\omega_0}\left(e^{-i\omega_0 t} - e^{i\omega_0 t}\right). \tag{7}$$

Using the Fourier transform, the $D^R(t)$ can be written in frequency form as

$$D^R(\omega) = \int_{-\infty}^{\infty} dt\, e^{i\omega_0 t} D^R(t), \tag{8}$$

$$D^R(\omega) = \frac{2\omega_0}{(\omega + i0^+)^2 - {\omega_0}^2}. \tag{9}$$

For the gamma point optical phonon mode with bare frequency $\omega_0$, the normalized free phonon propagator is written as [4]

$$[\widetilde{D}^R(\omega)]^{-1} = (\omega + \iota 0^+)^2 - {\omega_0}^2, \tag{10}$$

where, $\widetilde{D}^R(\omega) = \frac{D^R(\omega)}{2\omega_0}$. Now, considering the interaction of phonons with other degrees of freedom, the modified phonon propagator via Dyson's equation can be written as [4]

$$[\widetilde{D}^R(\omega)]^{-1} = (\omega + \iota 0^+)^2 - {\omega_0}^2 - 2\omega_0 \Pi(\omega, T), \tag{11}$$

here, $\Pi(\omega, T)$ is the phonon self-energy, which includes both the renormalized phonon frequency (real part) and phonon linewidth (imaginary part) due to both phonon-phonon and electron-phonon interactions.

Using the pole expansion, the complex pole of $\widetilde{D}^R(\omega)$ is expressed in terms of frequency shift and absolute linewidth as

$$\widetilde{\omega} = \omega_0 + \Delta\omega(T) - i\frac{\Gamma(T)}{2}. \tag{12}$$

Since $\widetilde{\omega}$ satisfy the Eq. (10), we have

$$\widetilde{\omega}^2 - {\omega_0}^2 - 2\omega_0 \varPi(\omega_0, T) = 0, \tag{13}$$

which can be rewritten as

$$\left(\omega_0 + \Delta\omega(T) - i\frac{\Gamma(T)}{2}\right)^2 - {\omega_0}^2 - 2\omega_0 \Pi(\omega_0, T) = 0. \tag{14}$$

From the expression in Eq. (14), the real and imaginary part of the phonon self-energy corresponds to the frequency shift and absolute linewidth as follows

$$\varDelta\omega(T) = \varPi'(\omega_0, T), \qquad \varGamma(T) = -2\varPi''(\omega_0, T). \tag{15}$$

In absence of electron-phonon interactions, only phonon-phonon interactions determine the phonon self-energy. The lowest-order anharmonic contribution originates from the cubic term, which couples one optical phonon to two acoustic phonons through three-phonon interactions.

$$H_3 = \frac{1}{3!}\sum_{ll'l''}\sum_{\alpha\beta\gamma} \Phi_{\alpha\beta\gamma}(ll'l'') u_{l\alpha} u_{l'\beta} u_{l''\gamma}\ , \tag{16}$$

where, $u_{l\alpha}$ is the displacement of atom in unit cell $l$ along the Cartesian direction $\alpha$, and $\Phi_{\alpha\beta\gamma}(ll'l'')$ are the third-order force constants.

After Fourier transformation and expansion in phonon eigenvectors, Eq. (16) can be written in terms of phonon operators as:

$$H_3 = \sum_{1,2,3} V(1,2,3)\,(a_1 + a_1^{\dagger})(a_2 + a_2^{\dagger})(a_3 + a_3^{\dagger})\,, \tag{17}$$

with an implicit crystal-momentum conservation, $q_1 + q_2 + q_3 = G$, where $G$ is a reciprocal lattice vector.

Klemens' decay model gives the anharmonic phonon self-energy, which involves a zone center phonon $(0, \omega_0)$ that decays into two acoustic phonons of equal energy and opposite momenta [5].

$$(0, \omega_0) \rightarrow (\vec{q}, \omega_q) + (-\vec{q}, \omega_{-q}), \tag{18}$$

where, $\omega_q = \omega_{-q} = \frac{\omega_0}{2}$

According to Fermi's golden rule applied to $H_3$, the imaginary part of the self-energy for the phonon decay process is expressed as [6]

$$\Gamma(\omega_0, T) = \frac{\pi}{\hbar^2} \sum_q \left|V_{0,q,-q}\right|^2 \delta(\omega_0 - \omega_q - \omega_{-q})(1 + n(\omega_q) + n(\omega_{-q}), \tag{19}$$

here, $V_{0,q,-q}$ is the three-phonon anharmonic coupling coefficient and $n(\omega_q)$, $n(\omega_{-q})$ are Bose Einstein occupation number for acoustic phonons with frequency $\omega_q$ and $\omega_{-q}$ respectively $(n(\omega) = 1/(e^{\hbar\omega/K_B T} - 1))$. Under the symmetric decay approximation, the Bose factor simplifies to $1 + 2\mathrm{n}(\omega_0/2)$.

Approximating the matrix element $\left|V_{0,q,-q}\right|^2$ and the relevant phonon density of states as slowly varying around the decay point, the sum over $q$ yields a temperature independent constant. Thus, the Klemens linewidth can be written as

$$\Gamma(T) = \Gamma_0 + D\left(\frac{\omega_0}{2}\right)\left[1 + 2n\left(\frac{\omega_0}{2}\right)\right], \tag{20}$$

where, $\Gamma_0$ is the temperature-independent residual of the linewidth (e.g. due to defects or inhomogeneity) and $D > 0$ is the anharmonic coupling constant.

The Kramers Kronig transformation [6] relates the phonon linewidth and frequency shift given in Eq. (15) as

$$\Delta\omega(\omega_0, T) = \frac{1}{\pi}\mathcal{P}\int_0^{\Lambda} \frac{\omega'\Gamma(\omega', T)}{{\omega_0}^2 - {\omega'}^2}\,d\omega', \tag{21}$$

here, $\Lambda$ is a high frequency cutoff and $\mathcal{P}$ denotes the Cauchy principal value of the integral.

For a general phonon frequency $\omega'$, the phonon linewidth arising from the symmetric decay channel can be written in the form

$$\Gamma(T) = D\left(\frac{\omega'}{2}\right)\left[1 + 2n\left(\frac{\omega'}{2}\right)\right], \tag{22}$$

where $D\left(\frac{\omega'}{2}\right)$ is a slowly varying prefactor that incorporates both the matrix elements and the available decay phase space.

Substituting this expression into Eq. (21) gives

$$\Delta\omega(\omega_0,T) = \frac{1}{\pi}\mathcal{P}\int_0^{\Lambda}\frac{\omega' D\left(\frac{\omega'}{2}\right)\left[1+2n\left(\frac{\omega'}{2}\right)\right]}{{\omega_0}^2-{\omega'}^2}d\omega', \tag{23}$$

which is dominated by contributions from frequencies $\omega'$ near the bare phonon frequency $\omega_0$, where the self-energy has its largest weight. Assuming that $D\left(\frac{\omega'}{2}\right)$ varies slowly near $\omega_0$ and that the decay phase space is narrow, we may approximate

$$D\left(\frac{\omega'}{2}\right) \approx D\left(\frac{\omega_0}{2}\right),\ n\left(\frac{\omega'}{2}\right) \approx n\left(\frac{\omega_0}{2}\right). \tag{24}$$

Under these approximations, the temperature-dependent factors can be extracted from the integral, leaving a principal-value integral that is independent of temperature and can be evaluated analytically as follows

$$\Delta\omega(\omega_0,T) \approx \left[1+2n\left(\frac{\omega_0}{2}\right)\right]\frac{D\left(\frac{\omega_0}{2}\right)}{\pi}\mathcal{P}\int_0^{\Lambda}\frac{\omega'}{{\omega_0}^2-{\omega'}^2}d\omega', \tag{25}$$

$$\mathcal{P}\int_0^{\Lambda}\frac{\omega'}{{\omega_0}^2-{\omega'}^2}d\omega' = -\frac{1}{2}\ln\left[\left(\frac{\Lambda}{\omega_0}\right)^2-1\right] \tag{26}$$

The integral in Eq. (26) yields a negative contribution in the limit $\Lambda \gg \omega_0$. By defining a positive constant $C$,

$$C = \frac{D\left(\frac{\omega_0}{2}\right)}{2\pi}\ln\left[\left(\frac{\Lambda}{\omega_0}\right)^2-1\right] \tag{27}$$

the expression reduces to the Klemens form for the phonon frequency shift

$$\Delta\omega(T) = \omega(T)-\omega_0 = -C\left[1+2\mathrm{n}\left(\frac{\omega_0}{2}\right)\right], \tag{28}$$

$$\omega(T) = \omega_0 - C\left[1+2\mathrm{n}\left(\frac{\omega_0}{2}\right)\right], \tag{29}$$

where, $C > 0$ is the anharmonic coupling constant, which is proportional to $D$, as given in Eq. (27).

### 3.2 Electron-phonon contribution to the phonon self-energy

For a normal metal, the Hamiltonian due to the interaction between electrons and lattice vibrations is written as [4]

$$H_{e-ph} = \sum_{q,k} g_q\left(a_q + a^{\dagger}{}_{-q}\right)c^{\dagger}{}_{k+q}c_k, \tag{30}$$

here, $c^{\dagger}{}_k(c_k)$ are creation(annihilation) operators for electrons with momentum k and $g_q$ is the electron-phonon coupling strength. The renormalized phonon frequency and linewidth arising from coupling are included in the modified phonon self-energy, which is expressed as

$$\Pi = \Pi_{ph-ph} + \Pi_{e-ph}, \tag{31}$$

where, $\Pi_{e-ph}$ is the self-energy from the electron-phonon contribution.

In the Raman limit $q \backsimeq 0$, the dynamic (non-adiabatic) electron phonon (e-ph) contribution to the phonon self-energy is given as follows [7]:

$$\Pi_{e-ph}(\omega_0,T) = \frac{2}{\hbar}\frac{1}{N_k}\sum_{k}\sum_{m\neq n}|g_{mn}(k)|^2[f_{mk}(T)-f_{nk}(T)]\left[\frac{1}{\Delta\varepsilon_{mnk}-\hbar(\omega_0+i\eta)}-\frac{1}{\Delta\varepsilon_{mnk}}\right]. \quad (32)$$

Here, a factor of 2 accounts for spin degeneracy, and the first sum is performed over $N_k$ electron wave vectors $k$. The second sum run over all the Kohn-Sham states, with Fermi-Dirac occupations $f_{nk} = f(\varepsilon_{nk}, \mathrm{T})$, where $\varepsilon_{nk}$ are the Kohn-Sham energies and $m, n$ are two electronic band indices. $\Delta\varepsilon_{mnk} = \varepsilon_{mk} - \varepsilon_{nk}$. The quantities $g_{mn}(k)$ denote the electron-phonon matrix elements, and $\eta$ is a real positive infinitesimal.

Using the identity

$$\frac{1}{x-i\eta} = \mathcal{P}\left(\frac{1}{x}\right) + i\pi\delta(x), \quad (33)$$

the real and imaginary parts of Eq. (32) are written as

$$\Pi''_{e-ph}(\omega_0,T) = \frac{2}{\hbar}\frac{1}{N_k}\sum_{k,m\neq n}|g_{mn}(k)|^2[f_{mk}(T)-f_{nk}(T)]\ \delta(\Delta\varepsilon_{mnk}-\hbar\omega_0) \quad (34)$$

$$\Pi'_{e-ph}(\omega_0,T) = \frac{2}{\hbar}\frac{1}{N_k}\sum_{k,m\neq n}|g_{mn}(k)|^2[f_{mk}(T)-f_{nk}(T)]\left[\mathcal{P}\frac{1}{\Delta\varepsilon_{mnk}-\hbar\omega_0}-\frac{1}{\Delta\varepsilon_{mnk}}\right] \quad (35)$$

The imaginary part $\Pi''_{e-ph}$ in Eq. (34) arises from real electron-hole pair creation with energy $\frac{\hbar\omega_0}{2}$ below (above) the Fermi level and contributes to the homogeneous phonon broadening. Because it contains a Dirac delta function enforcing strict energy conservation, the available phase space is strongly constrained. Thus, only electrons in the vicinity of the Fermi level contribute to the linewidth due to electron-phonon interactions. In good metals with a single partially filled band the Fermi energy is much larger than the phonon energy $\left(\mu \simeq E_F \gg \frac{\hbar\omega_0}{2}\right)$, the occupation difference $f_{mk}(T) - f_{nk}(T)$ varies only weakly with temperature, so the transition is effectively Pauli blocked. Consequently, this phase-space mechanism does not produce a pronounced smooth temperature dependence in $\Gamma(T)$ for $RuO_2$/$IrO_2$. Hence, Eq. (20), i.e., Klemens decay model for $\Gamma(T)$ still satisfies for metallic rutiles. This agrees with our experimental results that $\Delta\Gamma$ does not have a clear trend for metallic and insulating rutiles.

Eq. (35) expresses the phonon frequency shift as the real part of the electron-phonon self-energy written as a Brillouin-zone sum. From Eq. (35), all temperature dependence enters exclusively through the Fermi-Dirac occupation factors defined in Eq. (36); the electron-phonon matrix elements and the energy denominators are temperature independent.

For any electronic state with energy $\varepsilon$ the Fermi-Dirac distribution function is defined as follows

$$f(\varepsilon,T) = \frac{1}{e^{(\varepsilon-E_F)/K_BT}+1} \quad (36)$$

In Eq. (37), the temperature-independent contribution is separated explicitly, isolating the part of the self-energy that varies with temperature. The remaining temperature-dependent term is then rewritten in Eq. (40) as an integral over electronic energies. In this step, the ($k$)-dependent quantities, such as bandstructure, matrix elements, and energy denominators are collected into a smooth function $\mathcal{K}(\varepsilon;\omega_0)$, while the temperature dependence remains solely in the Fermi-Dirac distribution $f(\varepsilon, T)$.

$$\Delta\omega_{\text{e-ph}}(\omega_0, T) = \frac{2}{\hbar}\frac{1}{N_k}\sum_{k,m\neq n}|g_{mn}(k)|^2[f_{mk}(T) - f_{nk}(T)] \left[\mathcal{P}\frac{1}{\Delta\varepsilon_{mnk} - \hbar\omega_0}\right] + \text{T-independent constant.} \quad (37)$$

$$\Delta\omega_{\text{e-ph}}(\omega_0, T) = \int_{-\infty}^{\infty}\mathcal{K}(\varepsilon;\omega_0)f(\varepsilon,T)d\varepsilon + \text{T-independent constant.} \quad (38)$$

The essential observation is that in a metal, only electronic states within an energy window of order $k_B T$ around the Fermi level change occupation with temperature. States far below or far above $E_F$ remain fully occupied or empty and therefore do not contribute to the temperature variation of the integral in Eq. (38). Consequently, the temperature dependence of Eqs. (37)-(38) is governed entirely by the behaviour of the smooth kernel $(\mathcal{K}(\varepsilon;\omega_0))$ in the vicinity of $\varepsilon = E - E_F = 0$.

Since $RuO_2$ and $IrO_2$ satisfy $(K_B T \ll E_F)$ over the full experimental temperature range, the integral in Eq. (38) can be expanded systematically around $(\varepsilon = 0)$ using the Sommerfeld expansion. This expansion is a controlled mathematical procedure for integrals involving the Fermi-Dirac function in the low-temperature limit and does not introduce any additional physical assumptions.

Applying the Sommerfeld expansion [8] to Eq. (38) yields Eq. (39), which consists of a temperature-independent term plus a leading correction proportional to $T^2$, determined by the derivative $\mathcal{K}'(0;\omega_0)$. This term produces the additional smooth $T^2$ contribution to the phonon frequency shift observed in the metallic compounds.

$$\int_{-\infty}^{\infty}\mathcal{K}(\varepsilon;\omega_0)f(\varepsilon,T)d\varepsilon = \int_{-\infty}^{0}\mathcal{K}(\varepsilon;\omega_0)d\varepsilon + \frac{\pi^2}{6}(K_B T)^2\mathcal{K}'(0;\omega_0) + O(T^4) \quad (39)$$

Applying this result to Eq. (38), the frequency shift is expressed as

$$\Delta\omega_{\text{e-ph}}(T) = \Delta\omega_{\text{e-ph}}(0) + KT^2 \quad (40)$$

with $K = \frac{\pi^2}{6}{K_B}^2\mathcal{K}'(0;\omega_0)$.

For metallic $RuO_2$ and $IrO_2$, an additional electronic contribution modifies the frequency shift. By absorbing the temperature-independent part of the electron-phonon renormalization $\Delta\omega_{\text{e-ph}}(0)$ into an effective bare frequency $\omega_0^*$, we get $\omega(T)$ as

$$\omega(T) = \omega_0^* - C\left[1 + 2\text{n}\left(\frac{\omega_0^*}{2}\right)\right] + KT^2, \quad (41)$$

where, the $KT^2$ term captures the low-temperature electronic correction arising from electron-phonon interactions and the sign of $K$ depends on the electronic bandstructure of the material. Thus, by replacing $\omega_0$ with $\omega_0^*$ in Eq. (20), the $\Gamma(T)$ for $RuO_2$/$IrO_2$ can be written as

$$\Gamma(T) = \Gamma_0 + D\left(\frac{\omega_0^*}{2}\right)\left[1 + 2n\left(\frac{\omega_0^*}{2}\right)\right] \quad (42)$$

### 3.3 Explicit kernel in a simple band model

The smooth kernel from the real part of electron-phonon self-energy is written as

$$\mathcal{K}(\varepsilon;\omega_0) = \frac{2}{\hbar}\frac{1}{N_k}\sum_{k,m\neq n}|g_{mn}(k)|^2[\delta(\varepsilon - \varepsilon_{mk}) - \delta(\varepsilon - \varepsilon_{nk})]\left[\mathcal{P}\frac{1}{\Delta\varepsilon_{mnk} - \hbar\omega_0}\right] \quad (43)$$

To obtain an explicit form of the kernel $\mathcal{K}(\varepsilon;\omega_0)$ within a simple framework, we consider a minimal two-band model describing vertical interband transitions in the $k\to 0$ limit. In this model, the electronic dispersions for the two bands are approximated as parabolic,

$$\varepsilon_{1k}=a_1+\frac{k^2}{2m_1},\quad \varepsilon_{2k}=a_2+\frac{k^2}{2m_2}, \tag{44}$$

as shown in FIG. 3. Here, $a_1<0$ and $m_1>0$, so that band 1 crosses the Fermi level ($E_F=0$). Assuming $\varepsilon_{2k}>k_BT$, implemented via $a_2>k_BT$ and $m_2>0$, ensures that band-2 lies well above the Fermi level and remains essentially unoccupied. The Fermi wavevector is therefore defined by $\varepsilon_{1k_F}=0$, yielding $k_F^2=-2m_1a_1$. Within this model, we approximate the momentum integration by extending $k$ up to $k_m>k_F$, without considering the Brillouin zone boundaries.

By assuming a constant electron-phonon coupling within the relevant low-energy range around the Fermi level, it can be approximated as

$$|g_{mn}(k)|^2\approx\ g^2 \tag{45}$$

Under these assumptions, the kernel in Eq. (43) can be expressed as a principal value integral over momentum,

$$\mathcal{K}(\varepsilon;\omega_0)\propto\mathcal{P}\int_0^{k_m}k^2\,dk\left[\frac{\delta(\varepsilon-\varepsilon_{1k})-\delta(\varepsilon-\varepsilon_{2k})}{\varepsilon_{1k}-\varepsilon_{2k}-\hbar\omega_0}+\frac{\delta(\varepsilon-\varepsilon_{2k})-\delta(\varepsilon-\varepsilon_{1k})}{\varepsilon_{2k}-\varepsilon_{1k}-\hbar\omega_0}\right]. \tag{46}$$

Since band-2 lies well above the Fermi level, $\varepsilon_{2k}>k_BT$, the occupation of band-2 is negligible, which yields $\delta(\varepsilon-\varepsilon_{2k})=0$ in the energy range $-k_BT<\varepsilon<k_BT$. The expression then simplifies to

$$\mathcal{K}(\varepsilon;\omega_0)\propto\mathcal{P}\int_0^{k_m}k^2\,dk\,\delta(\varepsilon-\varepsilon_{1k})\left[\frac{1}{\varepsilon_{1k}-\varepsilon_{2k}-\hbar\omega_0}-\frac{1}{\varepsilon_{2k}-\varepsilon_{1k}-\hbar\omega_0}\right]. \tag{47}$$

Introducing the interband energy difference $\Delta\varepsilon_k=\varepsilon_{2k}-\varepsilon_{1k}$, the square bracket term simplifies to $-2\Delta\varepsilon_k/[\Delta\varepsilon_k^2-(\hbar\omega_0)^2]$. The remaining integral is evaluated using the identity for the delta function, with $\varepsilon_{1k}=a_1+k^2/2m_1$, which fixes $k^2=2m_1(\varepsilon-a_1)$ and yields $\int k^2dk\,\delta(\varepsilon-\varepsilon_{1k})\propto m_1(\varepsilon-a_1)$.

Thus, Eq. (47) can be rewritten as

$$\mathcal{K}(\varepsilon;\omega_0)\propto\mathcal{P}\left[\frac{\Delta\varepsilon_k m_1(\varepsilon-a_1)}{\Delta\varepsilon_k^2-(\hbar\omega_0)^2}\right]. \tag{48}$$

where the interband energy separation is given by

$$\Delta\varepsilon_k=(a_2-a_1)+k^2\left(\frac{1}{2m_2}-\frac{1}{2m_1}\right)=\frac{(m_1-m_2)\varepsilon+(a_2m_2-a_1m_1)}{m_2}. \tag{49}$$

From Eq. (48), it follows that the kernel exhibits a strong enhancement when the denominator becomes small, i.e., when the resonance condition $\Delta\varepsilon_k\approx\hbar\omega_0$ is satisfied. In this regime, the derivative $\mathcal{K}'(0;\omega_0)$ is proportional to $\omega_0$, indicating that higher-frequency phonons induce a stronger renormalization of the electronic response. Furthermore, the kernel is significantly enhanced when the condition $|\,a_2-a_1m_1/m_2\,|\sim\hbar\omega_0$ is satisfied, which also corresponds to the resonance condition at the Fermi surface. Physically, this reflects a situation in which the phonon energy matches the interband energy separation, allowing for near-resonant virtual electronic transitions that strongly enhance the electron-phonon interaction. This highlights the importance of having multiple bands in close proximity in energy. In multiband systems, such as a $d$-band metals $RuO_2$ and $IrO_2$, several interband energy differences $\Delta\varepsilon_{ij}(k) = \varepsilon_{jk} - \varepsilon_{ik}$ may simultaneously satisfy the resonance condition ($\Delta\varepsilon_k \approx \hbar\omega_0$) for a given phonon energy. As a result, the total kernel becomes a sum over contributions from multiple band pairs, leading to an overall enhancement of the coupling strength.

Note that in the particle-hole symmetric limit [e.g. $a_1 = -a_2, m_1 = -m_2$ in Eq. (44)], $\varepsilon_{2k} = -\varepsilon_{1k}$ and the interband electron-phonon kernel becomes symmetric about the Fermi level, $\mathcal{K}(\varepsilon;\omega_0) = \mathcal{K}(-\varepsilon;\omega_0)$. Consequently, kernel-derivative $\mathcal{K}'(0;\omega_0)$=0, and the leading $T^2$ contribution obtained from the Sommerfeld expansion vanishes. However, in real multiband metals ($RuO_2$/$IrO_2$), particle-hole symmetry is broken due to asymmetries in the electronic band structure, leading generically to $\mathcal{K}'(0) \neq 0$ and hence $K \neq 0$. This leads to a finite $T^2$ contribution in phonon renormalization.

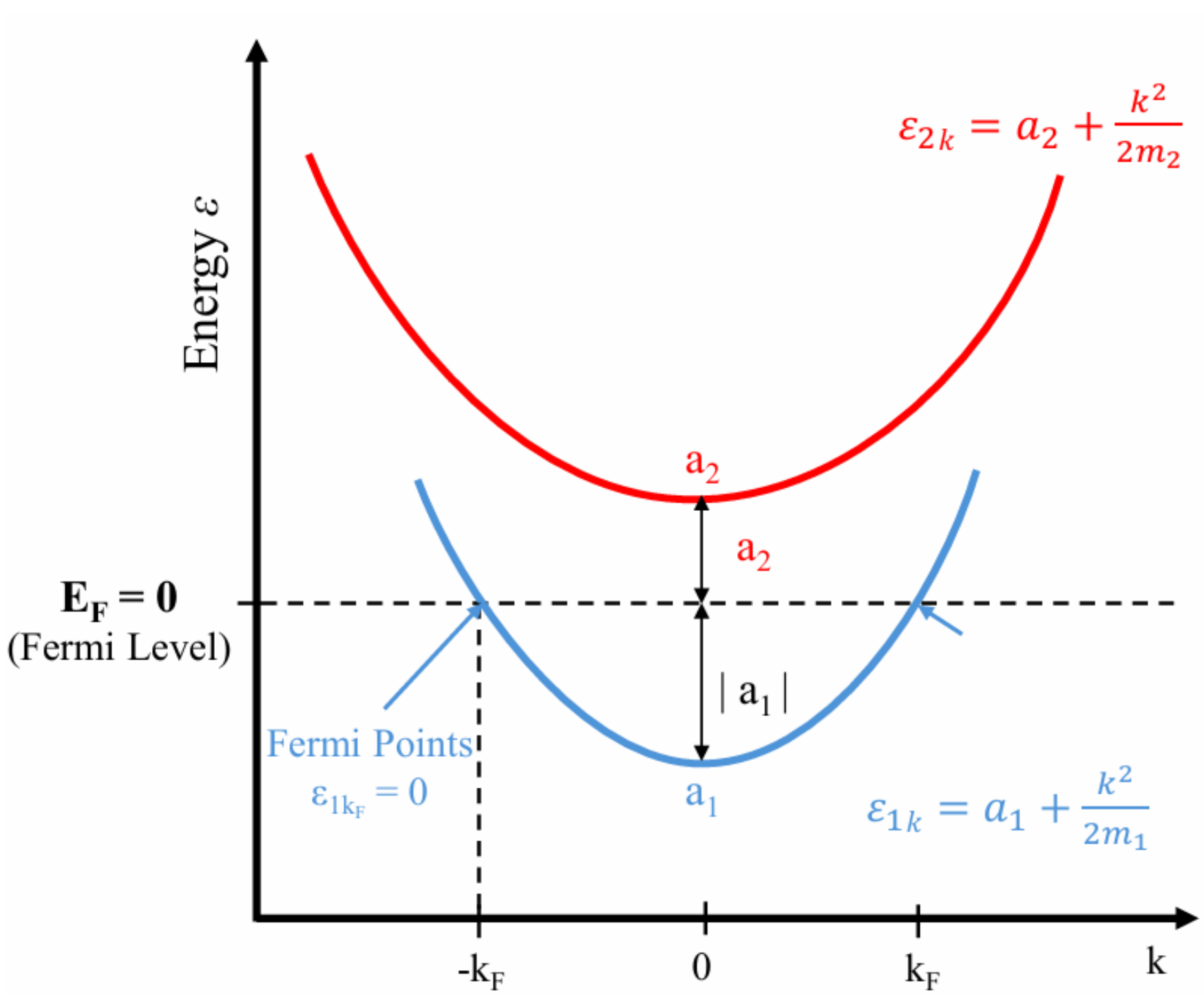

FIG. 3. Band structure near the $\Gamma$ point in a two-band model. The dispersions are parabolic, $\varepsilon_{ik} = a_i + k^2/2m_i$; the larger curvature of band 1 implies $m_1 < m_2$.

## References

[1] T. Berlijn, P. C. Snijders, O. Delaire, H. D. Zhou, T. A. Maier, H. B. Cao, S. X. Chi, M. Matsuda, Y. Wang, M.R. Koehler, P.R.C. Kent, and H.H. Weitering, Itinerant Antiferromagnetism in $RuO_2$, Phys. Rev. Lett. **118**, 077201 (2017).

[2] Y. S. Huang, S. S. Lin, C. R. Huang, M. C. Lee, T. E. Dann, and F. Z. Chien, Raman spectrum of $IrO_2$, Solid State Commun. **70**, 517 (1989).

[3] P. S. Peercy and B. Morosin, Pressure and Temperature Dependences of the Raman-Active Phonons in $SnO_2$, Phys. Rev. B **7**, 2779 (1973).

[4] A. B. Migdal, Interaction between electrons and lattice vibrations in a normal metal, J. Exptl. Theor. Phys. **34**, 1438 (1958).

[5] P. G. Klemens, Anharmonic decay of optical phonons, Phys. Rev. **148**, 845 (1966).

[6] J. Menéndez and M. Cardona, Temperature dependence of the first-order Raman scattering by phonons in Si, Ge, and α-Sn: Anharmonic effects, Phys. Rev. B **29**, 2051 (1984).

[7] M. Lazzeri and F. Mauri, Nonadiabatic Kohn anomaly in a doped graphene monolayer, Phys. Rev. Lett. **97**, 266407 (2006).

[8] N. W. Ashcroft and N. D. Mermin, Solid State Physics (Harcourt, Orlando, FL, 1976).